\documentclass[runningheads,orivec]{llncs}
\usepackage[T1]{fontenc}
\usepackage{graphicx}
\usepackage{booktabs}
\usepackage[caption=false]{subfig}
\usepackage{svg}
\usepackage{float}
\usepackage[most]{tcolorbox}
\usepackage{framed}
\svgsetup{
  inkscapeopt= { --export-latex --pdf-version=1.5 }, 
}
\usepackage{footnote}
\makesavenoteenv{figure}

\usepackage[hidelinks]{hyperref}
\usepackage{url}
\usepackage{orcidlink}

\begin{document}
\title{GeoOutageKG: A Multimodal Geospatiotemporal Knowledge Graph for Multiresolution Power Outage Analysis} 
\titlerunning{GeoOutageKG}
%

\author{Ethan Frakes\inst{1}\orcidlink{0009-0008-8869-5703}, Yinghui Wu\inst{2}\orcidlink{0000-0003-3991-5155}, Roger H. French\inst{2}\orcidlink{0000-0002-6162-0532}, Mengjie Li\inst{1}\orcidlink{0000-0002-9531-2474}
}
\authorrunning{E. Frakes et al.}
%
\institute{
  \inst{} University of Central Florida, Orlando, FL 32816, USA\\
  \email{\{ethan.frakes, mengjie.li\}@ucf.edu}
  \and
  \inst{} Case Western Reserve University, Cleveland, OH 44106, USA\\
  \email{\{yinghui.wu2, roger.french\}@case.edu}
}

%
\maketitle              
\begin{abstract}

    Detecting, analyzing, and predicting power outages is crucial for grid risk assessment and disaster mitigation. 
    Numerous outages occur each year, exacerbated by extreme weather events such as hurricanes. Existing outage data are typically reported at the county level, limiting their spatial resolution and making it difficult to capture localized patterns. However, it offers excellent temporal granularity.
    In contrast, nighttime light satellite image data provides significantly higher spatial resolution and enables a more comprehensive spatial depiction of outages, enhancing the accuracy of assessing the geographic extent and severity of power loss after disaster events. 
    However, these satellite data are only available on a daily basis.
    Integrating spatiotemporal visual and time-series data sources into a unified knowledge representation can substantially improve power outage detection, analysis, and predictive reasoning. 
    In this paper, we propose GeoOutageKG, a multimodal knowledge graph that integrates diverse data sources, including nighttime light satellite image data, high-resolution spatiotemporal power outage maps, and county-level timeseries outage reports in the U.S. 
    We describe our method for constructing GeoOutageKG by aligning source data with a developed ontology, GeoOutageOnto. Currently, GeoOutageKG includes over 10.6 million individual outage records spanning from 2014 to 2024, 300,000 NTL images spanning from 2012 to 2024, and 15,000 outage maps.
    GeoOutageKG is a novel, modular and reusable semantic resource that enables robust multimodal data integration. We demonstrate its use through multiresolution analysis of geospatiotemporal power outages.

    \textbf{OSF Repository:} \url{https://doi.org/10.17605/OSF.IO/QVD8B}
    
    \textbf{GitHub Repository:} \url{https://purl.org/geooutagekg}

    \textbf{Ontology Documentation:} \url{https://purl.org/geooutageonto}
    
\end{abstract}
%

\section{Introduction} 
    \label{sec1}
    
    The prediction and analysis of local power outages are critical tasks for assessing community energy vulnerability and disaster mitigation. 
    In 2024, Hurricanes Helene and Milton caused widespread destruction and power outages, with Hurricane Helene alone responsible for approximately 1.69 million outages in the state of Florida \cite{Hagen_Cangialosi_Chenard_Alaka_Delgado_2025}. 
    Power outage analysis is essential for understanding the scale and impact of such events, guiding emergency response, infrastructure recovery, and long-term grid resilience planning.
    
    Various methods \cite{abdelmalak_2023,aparcedo2024multimodalpoweroutageprediction} have been developed for power outage analysis, but most focused on a single data modality, such as tracking the number of customers without power or visualizing individual outage locations, without incorporating broader contextual information. 
    Abdelmalak et al. \cite{abdelmalak_2023} focused on developing a resilience assessment framework to evaluate the characteristics of extreme outage events at the state level in the United States. The framework is built to evaluate the power system performance based on the Environment for Analysis of Geo-Located Energy Information (EAGLE-I) dataset \cite{brelsford2024dataset,eaglei}.
    Aparcedo et al. \cite{aparcedo2024multimodalpoweroutageprediction} developed a visual-spatiotemporal graph neural network (VST-GNN) to create visual outage predictions and risk maps using nighttime light (NTL) satellite image data from NASA's Black Marble \cite{blackmarble} dataset. 
    Although multiple datasets exist for analyzing power outages, each has inherent limitations. 
    For example, existing outage data such as EAGLE-I are typically reported at the county level, which constrains spatial resolution and makes it challenging to capture localized outage patterns. 
    Nonetheless, these records offer excellent temporal granularity, with updates available as frequently as every 15 minutes. 
    In contrast, nighttime light (NTL) satellite imagery provides significantly higher spatial resolution and enables a more comprehensive spatial representation of outages.
    This in turn improves assessment of the geographic extent and severity of power loss following disaster events. 
    However, NTL data are typically available only on a daily basis, limiting their temporal responsiveness.
    Despite the availability of these diverse datasets, a comprehensive and integrated knowledge base that semantically unifies them is still lacking.
    Integrating these complementary spatiotemporal visual and time series data sources into a unified multimodal knowledge representation can substantially improve power outage detection, analysis, and predictive reasoning.
    
    To address these limitations, we propose GeoOutageKG, a multimodal knowledge graph that semantically integrates heterogeneous data sources to support power outage analysis at multiple spatiotemporal resolutions. 
    By aligning high-temporal-resolution outage records (e.g., EAGLE-I) with high-spatial-resolution satellite image data (e.g., NTL images from Black Marble), GeoOutageKG enables more granular, context-aware analysis of outage events. 
    The knowledge graph not only facilitates spatiotemporal correlation and reasoning across data modalities, but also supports extensibility for incorporating additional datasets.
    This can include meteorological records, infrastructure assets, and sociodemographic indicators. 
    This unified semantic representation lays the foundation for advanced use cases in grid resilience assessment, equitable disaster response, and predictive outage modeling.

    At the time of writing, GeoOutageKG includes over 10.6 million individual outage records spanning from 2014 to 2024, over 300,000 NTL images spanning from 2012 to 2025, and over 15,000 outage maps generated from the five most recent major hurricanes in Florida.
    GeoOutageKG's ontology and knowledge graph are implemented in OWL2 \cite{w3cOWL2Web2012} and represented using the Resource Description Framework (RDF) 1.2 Schema \cite{w3cRDF12Schema2024}, serialized in Turtle syntax \cite{beckett_2014}.

\section{Related Work} 
    \label{sec2}

    \subsection{Grid Performance Assessment} 
    \label{subsec2.1}   
    
    Although our work primarily targets geospatiotemporal outage analysis, other ontology-based approaches have been developed with a focus on grid performance assessment. 
    The Electric Power Fault Detection Ontology (EPFDO) represents a network of related ontologies designed to model key elements and parameters within the power grid, including $sensors$, $actuators$, $energy\ quality$, $energy\ failures$, and $geographic\ location$ \cite{mederos_2020}. 
    EPFDO builds upon several existing ontologies, including the Ontology for Energy Management Applications (OEMA) \cite{cuenca_2017}. 
    OEMA consolidates multiple domain-specific vocabularies into a unified schema for representing energy performance and contextual information. 
    The OEMA network itself is structured into eight modular ontologies: $infrastructure$, $energy\ and\ equipment$, $geographical$, $external\ factors$, $person\ and \ organisation$, $ontology\ network$, $smart\ grid\ stakeholders$, and $units$. 
    
    In addition to efforts focused on ontology construction, other works explored the use of ontologies to support the automation and validation of outage and grid performance analysis. 
    Huang et al. \cite{huang_2022} developed a method for automatically identifying the causes of blackouts by integrating Natural Langauge Processing (NLP) with a manually created ontology. 
    Their approach analyzes web-sourced textual data, such as news articles and online reports, to extract and classify information related to blackout events.
    Mahmoud et al. \cite{mahmoud_2020} developed an ontology-based maintencance tool for detecting power substation faults in distribution grids. 
    The ontology models substation configuration and component interdependencies, enabling substation failure prediction and generation of maintenance reports based on component relationships and environmental conditions. 

    \subsection{Knowledge Graphs and Ontologies for Satellite Data} 
    \label{subsec2.2}   
        
    Given the critical role of Earth Observation (EO) satellites in modern scientific research, many structured databases and ontologies have been developed to catalog satellite systems and their onboard sensors. 
    The World Meteorological Organization (WMO) maintains the Observing Systems Capability Analysis and Review Tool (OSCAR), a dataset focused on EO satellites and sensors with primary applications in meteorology and climatology \cite{wmo_2023}. 
    Similarly, the Union of Concerned Scientists (UCS) created the UCS Satellite Database (UCSSD), which includes metadata on 7,560 active satellites in various domains, including geo-electronics and telecommunications \cite{ucs_2023}.
    This database served as the foundation for the UCS Satellite Ontology (UCSSO), which semantically organizes satellite attributes, such as $name$, $launch\ vehicle$, $launch\ date$, $operator$, and $orbital\ parameters$, into machine-readable classes and properties \cite{rovetto_2017}. 
    However, UCSSO is limited to satellites currently in orbit and does not capture information on decommissioned or de-orbited assets.
    To address this gap, Lin et al. introduced GEOSatDB \cite{lin_2024}, a knowledge graph specifically focused on EO satellites and their mounted sensors. 
    GEOSatDB includes 2,340 satellites, both operational and retired, and 1,021 sensors, categorized into four primary ontological classes: $Satellite$, $Sensor$, $OperationalBand$ (indicating the frequency range of the sensor) and $Operation$ (describing the sensor function).
    
    \subsection{Geospatial Ontologies and Query Standards} 
    \label{subsec2.3}  
    
    Various ontologies and knowledge graphs have been developed to represent and query geospatial data. 
    A key standard in this space is GeoSPARQL \cite{battle_2012}, which enables the representation and querying of geospatial data on the Semantic Web. 
    GeoSPARQL extends SPARQL \cite{harris_2013} with geospatial query capabilities and defines an RDF-based vocabulary for encoding spatial objects, such as coordinates, geographic points, and named landmarks (e.g., “Washington Monument” or “Mount Rushmore”).

    Furthermore, analogously to GeoOutageKG, standards have been developed for broader spatiotemporal data querying, such as the Spatio-Temporal Asset Catalogs (STAC) specification \cite{newman_2023}. 
    Though not an OWL2 \cite{w3cOWL2Web2012} standard ontology, this schema allows for standardized retrieval of spatiotemporal data and is formatted in a JSON-derived STAC format.
    Other geospatial and spatiotemporal data methods such as TerraQ \cite{kefalidis_2025} and EarthQA \cite{punjani_2023} leverage Natural Language Processing (NLP) to perform Text-to-SPARQL querying of Earth Observation (EO) satellite archives. 
    EarthQA specifically expands upon the AI4Copernicus \cite{ai4copernicus} framework, which provides an AI-on-demand framework for querying EO data and providing AI tools and datasets for training and deploying EO AI models.

    \subsection{Multimodal and Image Ontologies} 
    \label{subsec2.4}
    While GeoOutageKG is designed as a multimodal knowledge graph specified for outage analysis, other multimodal ontologies and knowledge graphs have been developed for more general use cases.
    Products such as MMKG \cite{liu_2019} and Richpedia \cite{wang_2020} are designed as multimodal knowledge graphs that semantically link images with respective textual entities and data.
    MMKG consists of three knowledge graphs that build from DBpedia \cite{dbpedia}, YAGO \cite{suchanek_2007}, and Freebase \cite{bollacker_2008,bordes_2013}.
    Richpedia aims to link textual entities from Wikidata \cite{wikidata2025} to respective visual and image entities.

\section{GeoOutageKG: Geospatiotemporal Power Outage Knowledge Graph} 
    \label{sec3}

    \begin{figure}[t]
        \centering
        \includegraphics[scale=0.27]{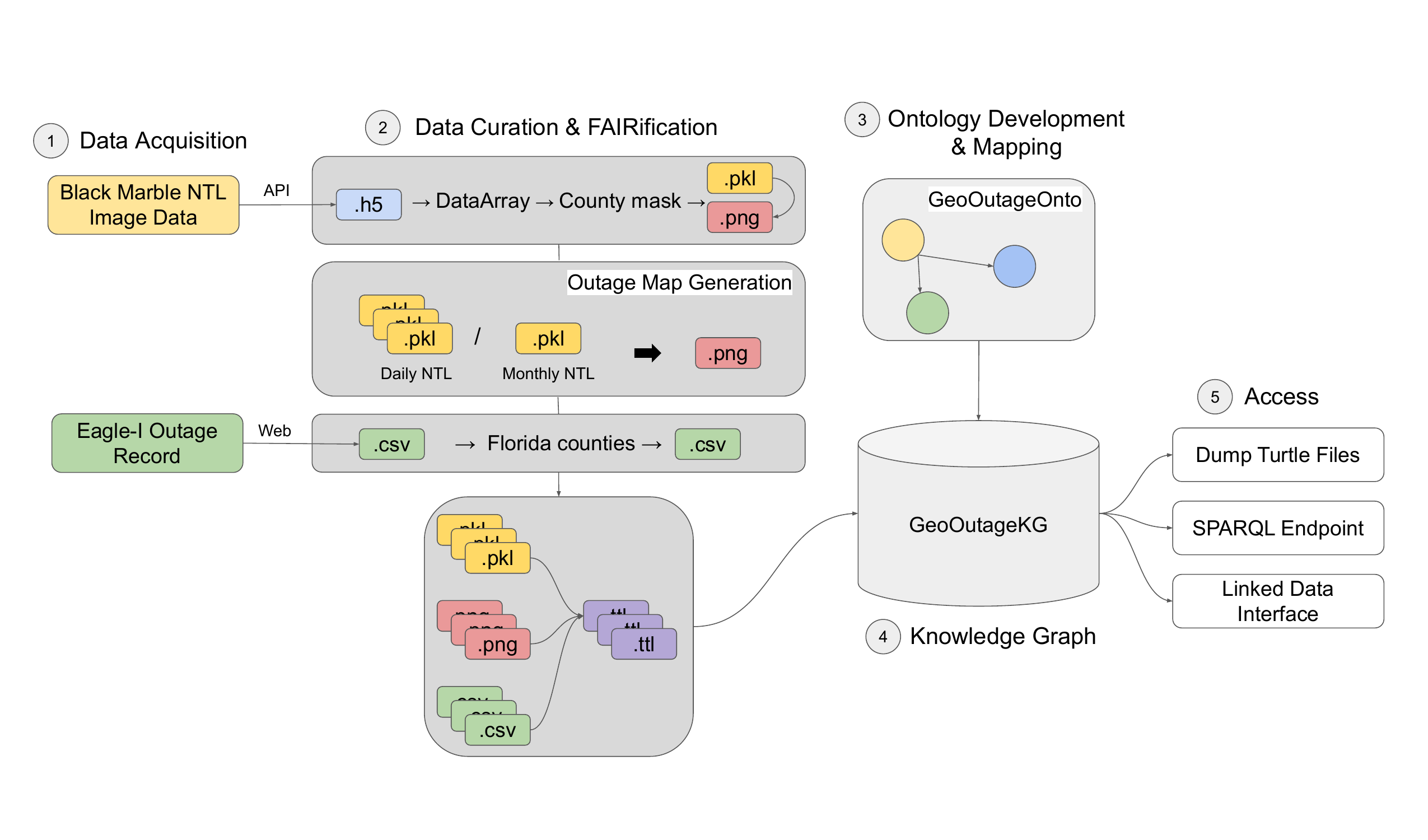}
        \caption{Continuously updated GeoOutageKG construction pipeline.}
        \label{fig:GeoOutageKGPipeline}
    \end{figure}

    In this section, we introduce our approach for the construction of the Geospatiotemporal Power Outage Knowledge Graph, GeoOutageKG. 
    Section \ref{subsec3.1} provides an overview of the construction pipeline and the key components.  
    Section \ref{subsec3.2} describes how we collect, filter and FAIRify the data.
    Section \ref{subsec3.3} presents the details on the ontology development. 
    
    \subsection{GeoOutageKG Construction Pipeline} 
    \label{subsec3.1}   
    
    Figure \ref{fig:GeoOutageKGPipeline} shows an overview of the GeoOutageKG construction pipeline.
    
    \textbf{Data Acquisition.} 
    This component retrieves and downloads raw data to integrate into GeoOutageKG. 
    For example, satellite image data, such as nighttime light (NTL) images, are accessible through the API of the Black Marble product suite \cite{blackmarble}.
    Outage records, a timeseries dataset, are accessible through the EAGLE-I dataset \cite{brelsford2024dataset,eaglei}.
    More details on the data we extract is found in Section \ref{subsec3.2}.

    \textbf{Data Curation and FAIRification.}
    Raw data can be filtered by time range and geographic location.
    For instance, in a case study focused on power outage analysis in Florida following the five most recent major hurricane events, nighttime light (NTL) satellite imagery is filtered and partitioned by county using geographic bounding boxes to cover all 67 counties in the state. 
    Additional analyses can be applied at this stage to enrich the dataset prior to integration. 
    For example, outage severity maps are generated by comparing NTL image data across pre- and post-event periods. 
    The curated dataset is then FAIRified, with the FAIRification process further explained in Section \ref{subsec3.2}.
    
    \textbf{Ontology Development and Mapping.} 
    This component develops the ontology and semantically maps the data source to its structure.
    We begin by analyzing the curated dataset to identify domain-relevant concepts, properties and relationships. 
    From this, we derive the core schema required to semantically model the data. 
    We then align the schema elements with existing terms from open, widely adopted ontologies to maximize reuse and interoperability. 
    New classes and properties are defined to cover domain specific concepts not represented in exiting vocabularies. 
    The resulting ontology, GeoOutageOnto, is further explained in Section \ref{subsec3.3}.
    
    \textbf{Knowledge Graph.}
    This step maps the curated data to GeoOutageOnto to generate the knowledge graph, GeoOutageKG. 
    We serialize the graph using the Turtle \cite{beckett_2014} file format. 
    
    \textbf{GeoOutageKG Access.}
    All class instances in GeoOutageKG are saved in Turtle \cite{beckett_2014} syntax. 
    All Turtle files and image data are openly available in our OSF repository\footnote[1]{\url{https://doi.org/10.17605/OSF.IO/QVD8B}}.
    GeoOutageKG is stored as a graph repository in GraphDB \cite{GraphDBGraphDB1102025}, which provides access via SPARQL \cite{harris_2013} endpoint for querying and integration. 
    Additionally, GeoOutageKG is the geospatial component and integrated in a broader ontology project, MDS-Onto \cite{rajamohan_2025}.
    MDS-Onto offers a linked data interface and API to support additional contributions of ontologies, data and metadata.

    \subsection{Acquiring, Curating and FAIRifying data} \label{subsec3.2}
    
\begin{figure}[!tb]
            \centering
            \begin{minipage}{0.135\linewidth}
                \includegraphics[width=\linewidth]{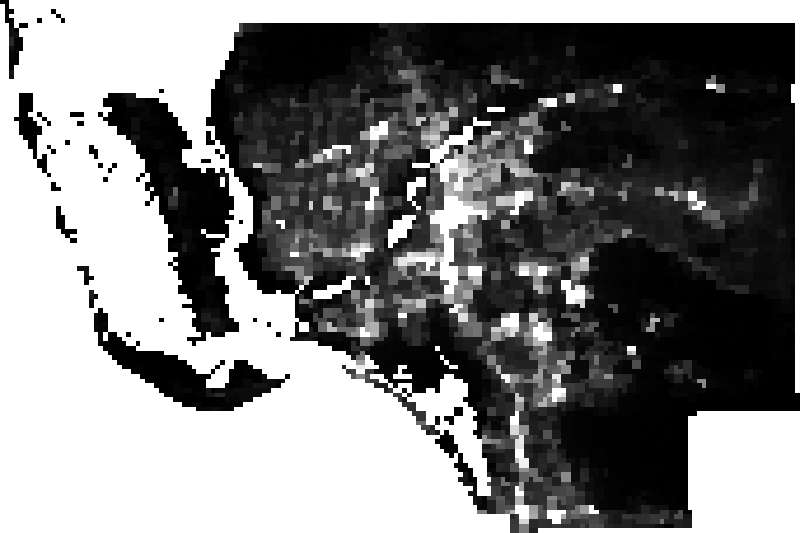}
            \end{minipage}
            \hfill
            \begin{minipage}{0.135\linewidth}
                \includegraphics[width=\linewidth]{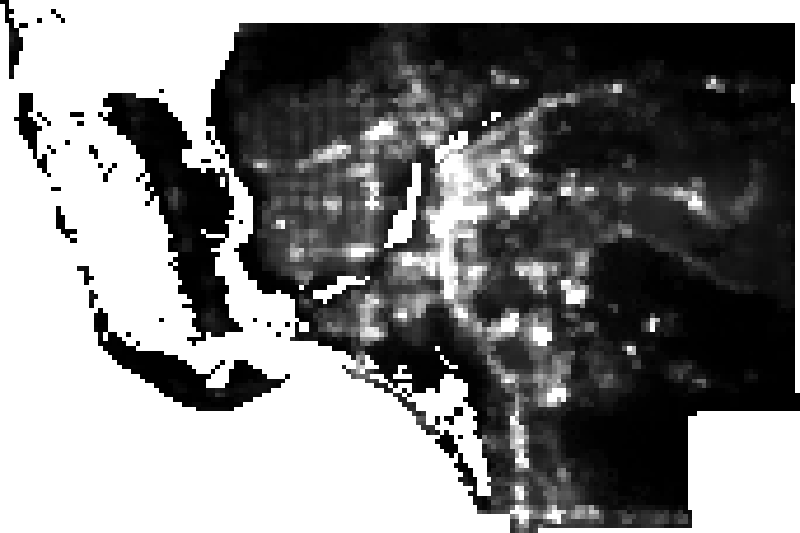}
            \end{minipage}
            \hfill
            \begin{minipage}{0.135\linewidth}
                \includegraphics[width=\linewidth]{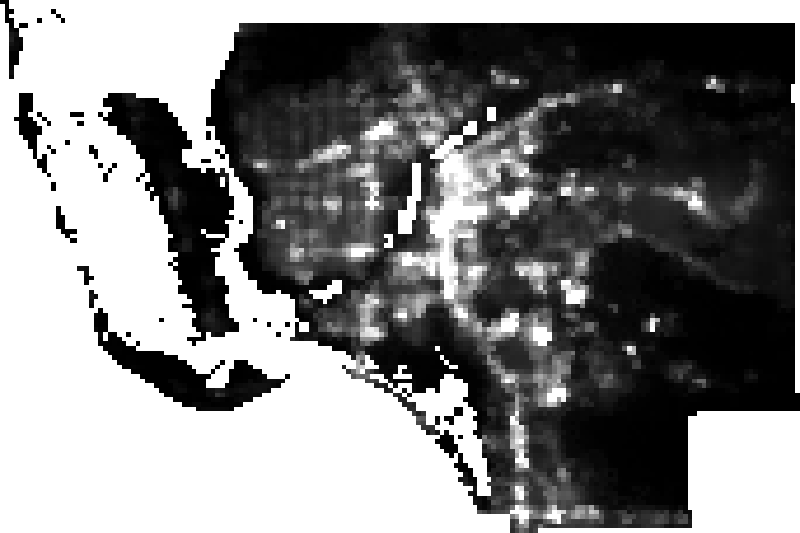}
            \end{minipage}
            \hfill
            \begin{minipage}{0.135\linewidth}
                \includegraphics[width=\linewidth]{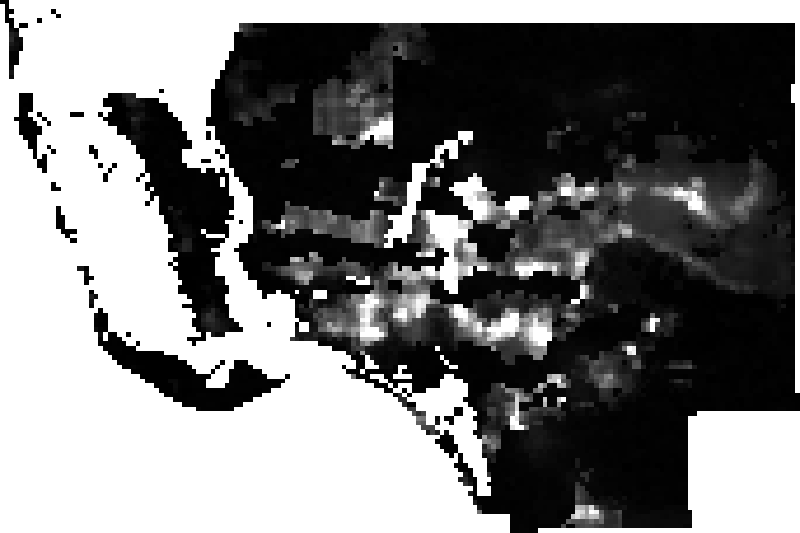}
            \end{minipage}
            \hfill
            \begin{minipage}{0.135\linewidth}
                \includegraphics[width=\linewidth]{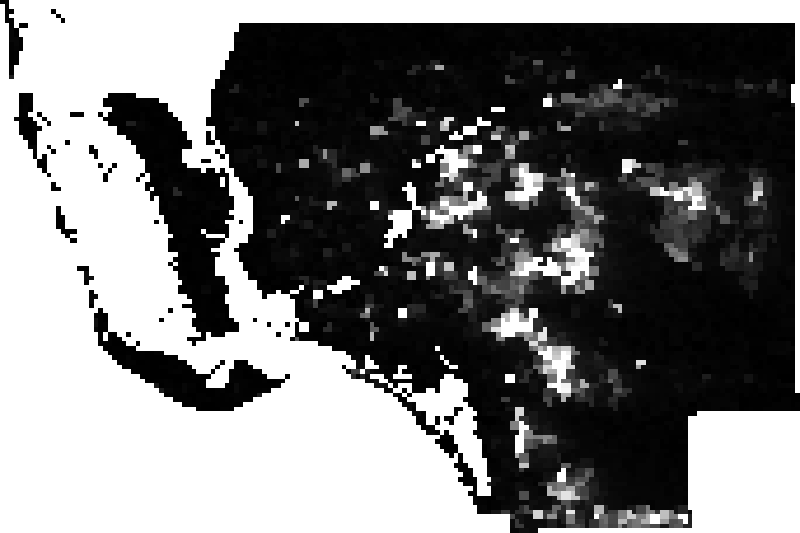}
            \end{minipage}
            \hfill
            \begin{minipage}{0.135\linewidth}
                \includegraphics[width=\linewidth]{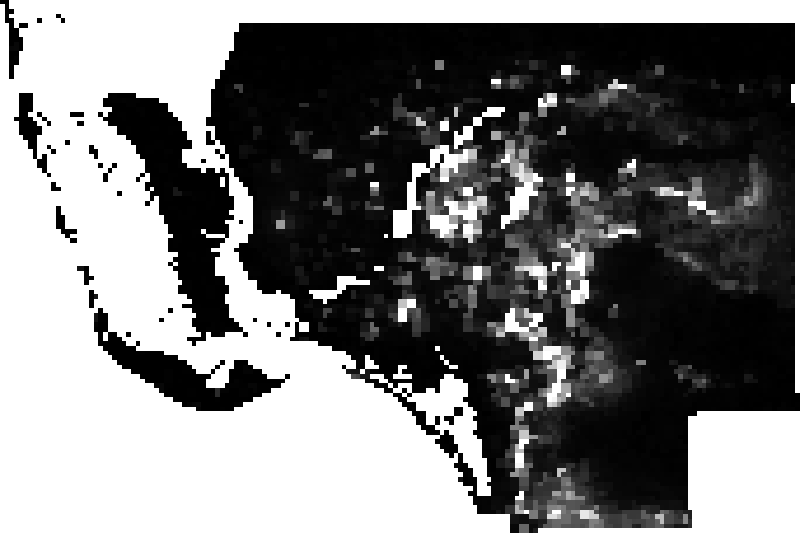}
            \end{minipage}
            \hfill
            \begin{minipage}{0.135\linewidth}
                \includegraphics[width=\linewidth]{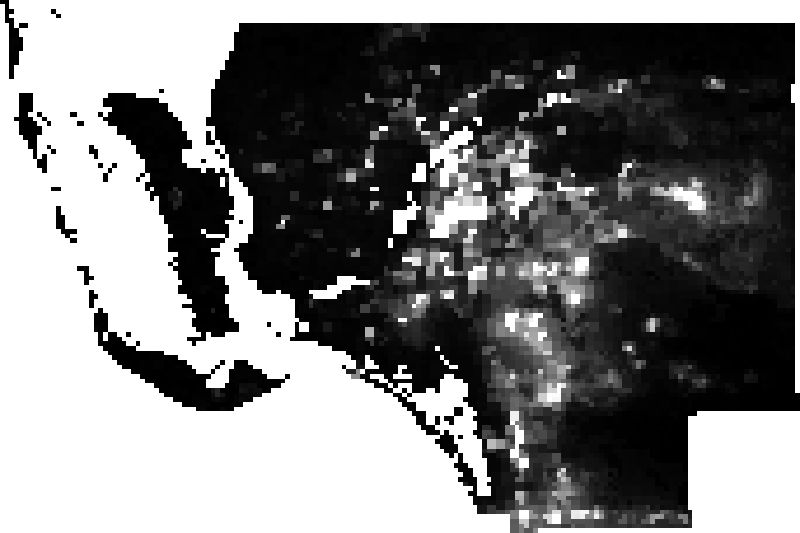}
            \end{minipage}
        
            \vspace{0.1cm}
        
            \begin{minipage}{0.135\linewidth}
                \includegraphics[width=\linewidth]{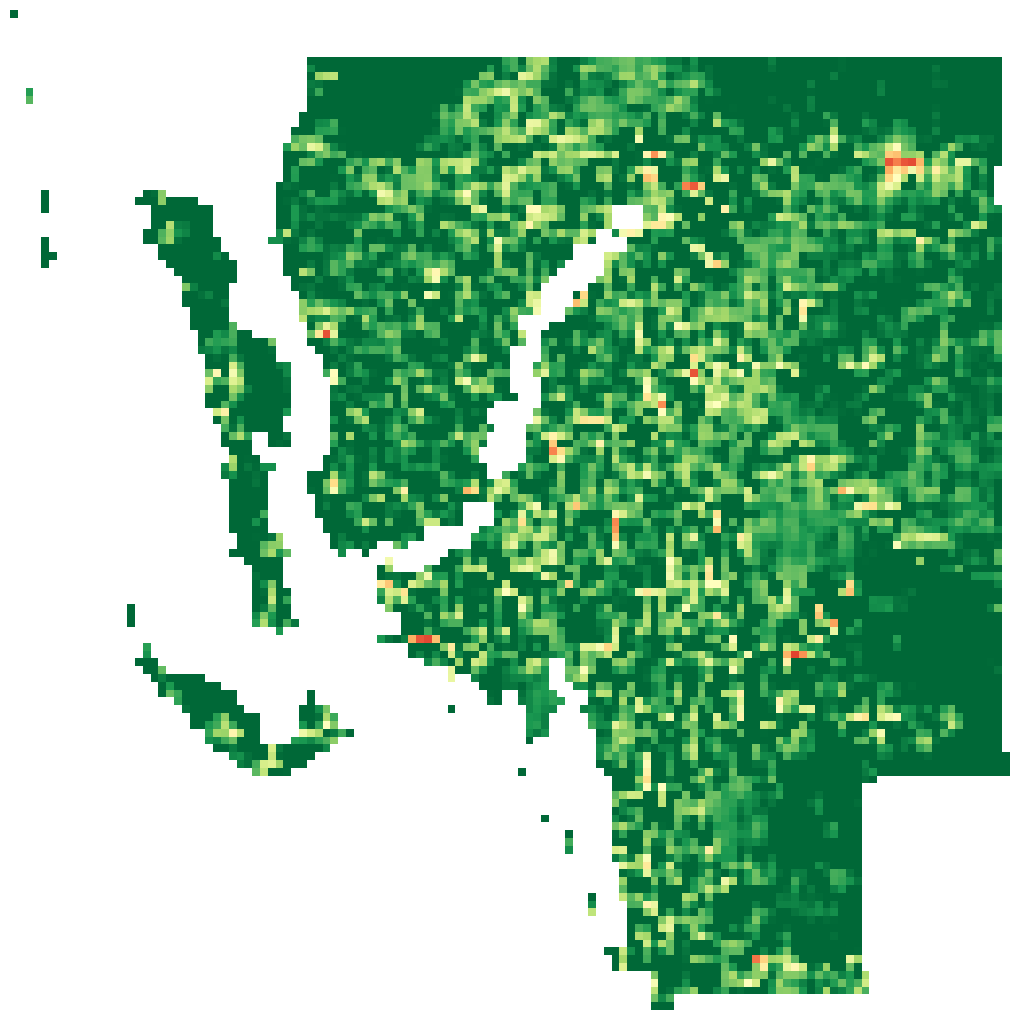}
            \end{minipage}
            \hfill
            \begin{minipage}{0.135\linewidth}
                \includegraphics[width=\linewidth]{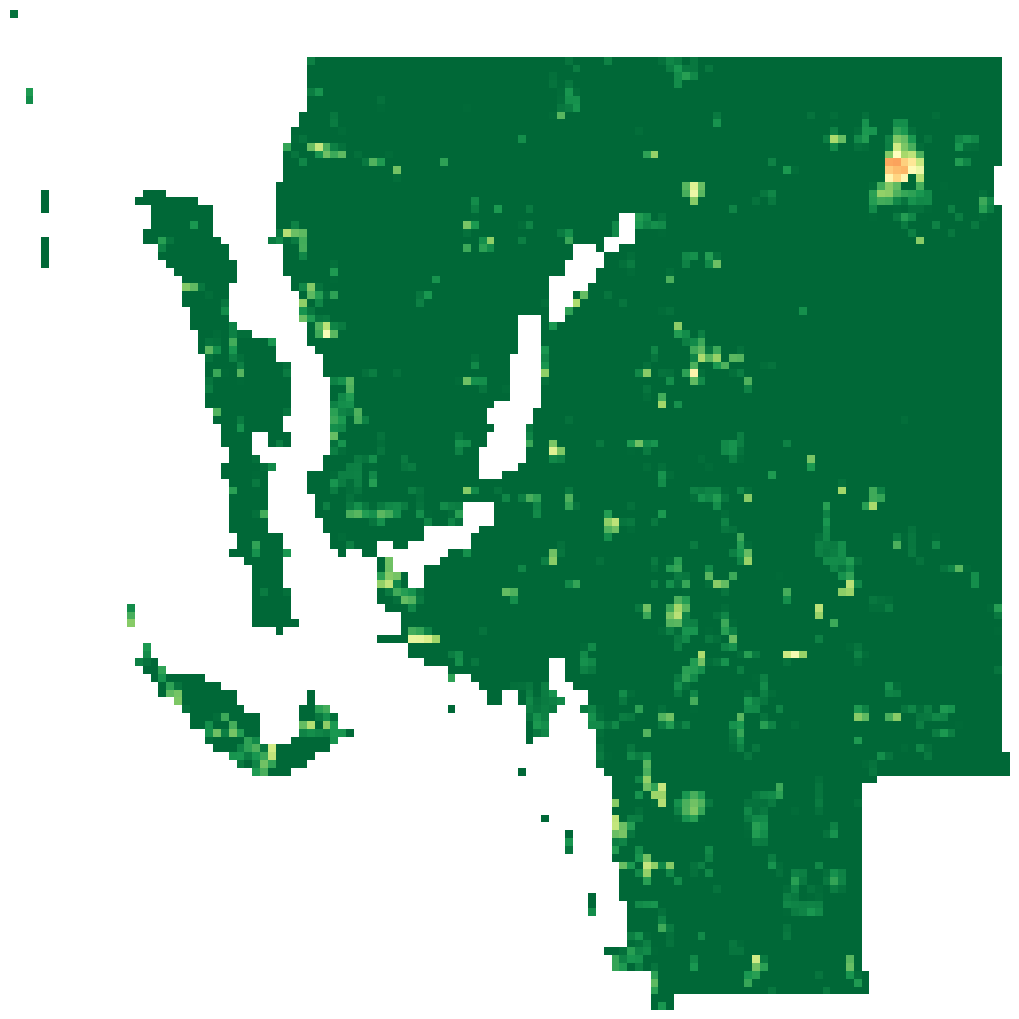}
            \end{minipage}
            \hfill
            \begin{minipage}{0.135\linewidth}
                \includegraphics[width=\linewidth]{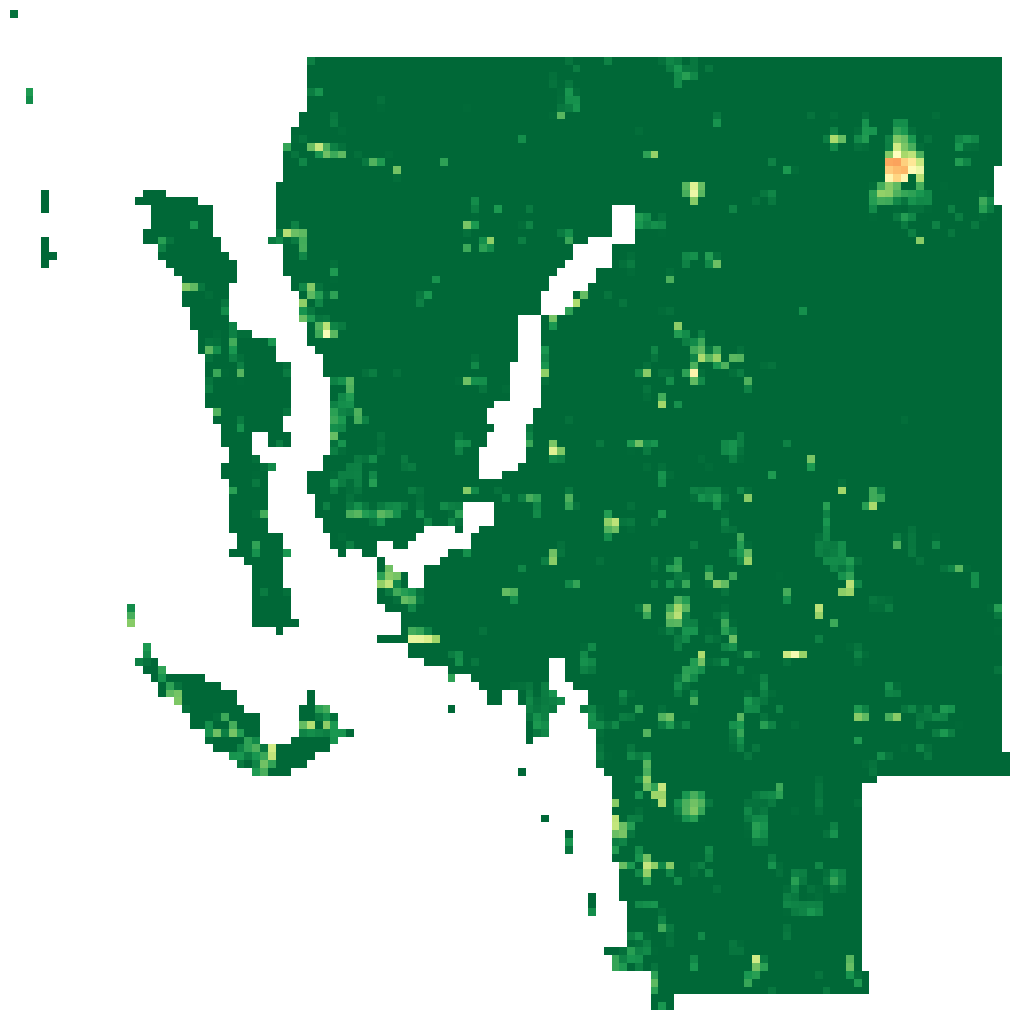}
            \end{minipage}
            \hfill
            \begin{minipage}{0.135\linewidth}
                \includegraphics[width=\linewidth]{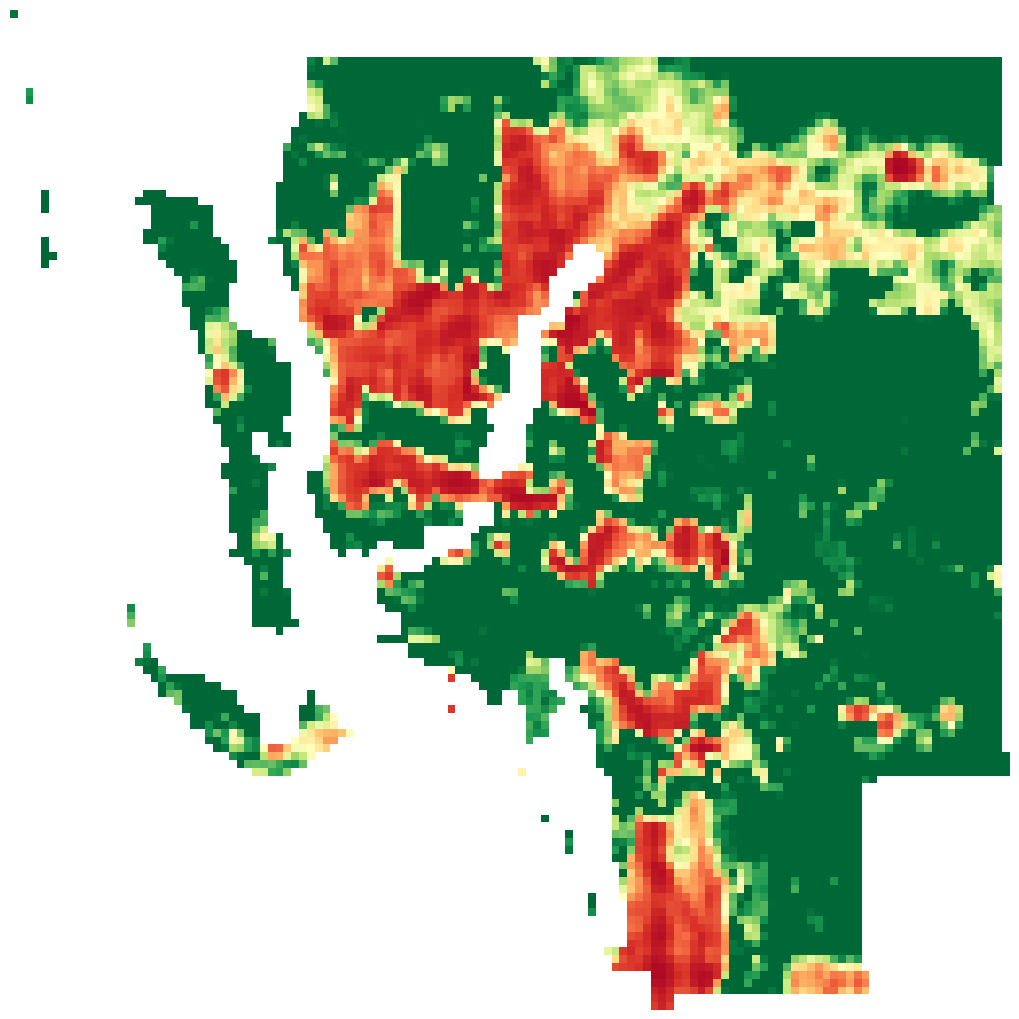}
            \end{minipage}
            \hfill
            \begin{minipage}{0.135\linewidth}
                \includegraphics[width=\linewidth]{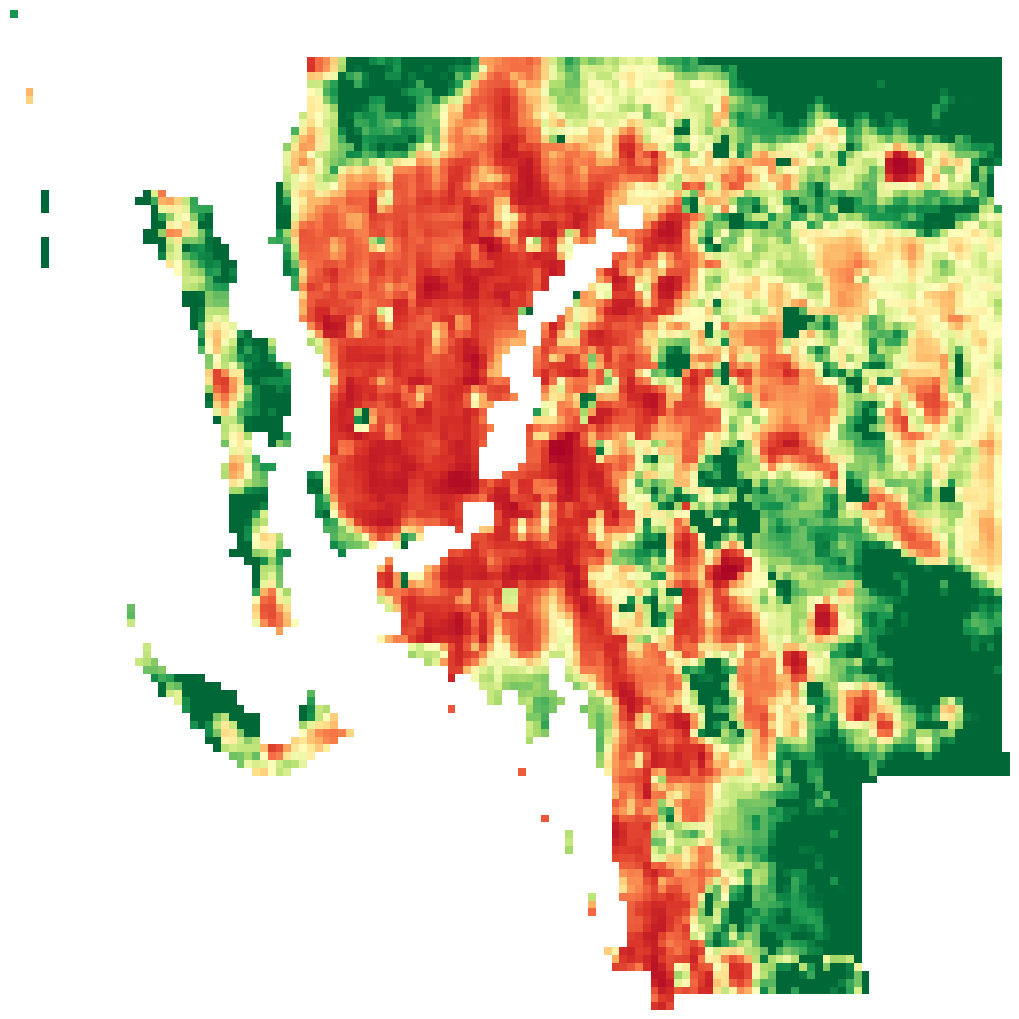}
            \end{minipage}
            \hfill
            \begin{minipage}{0.135\linewidth}
                \includegraphics[width=\linewidth]{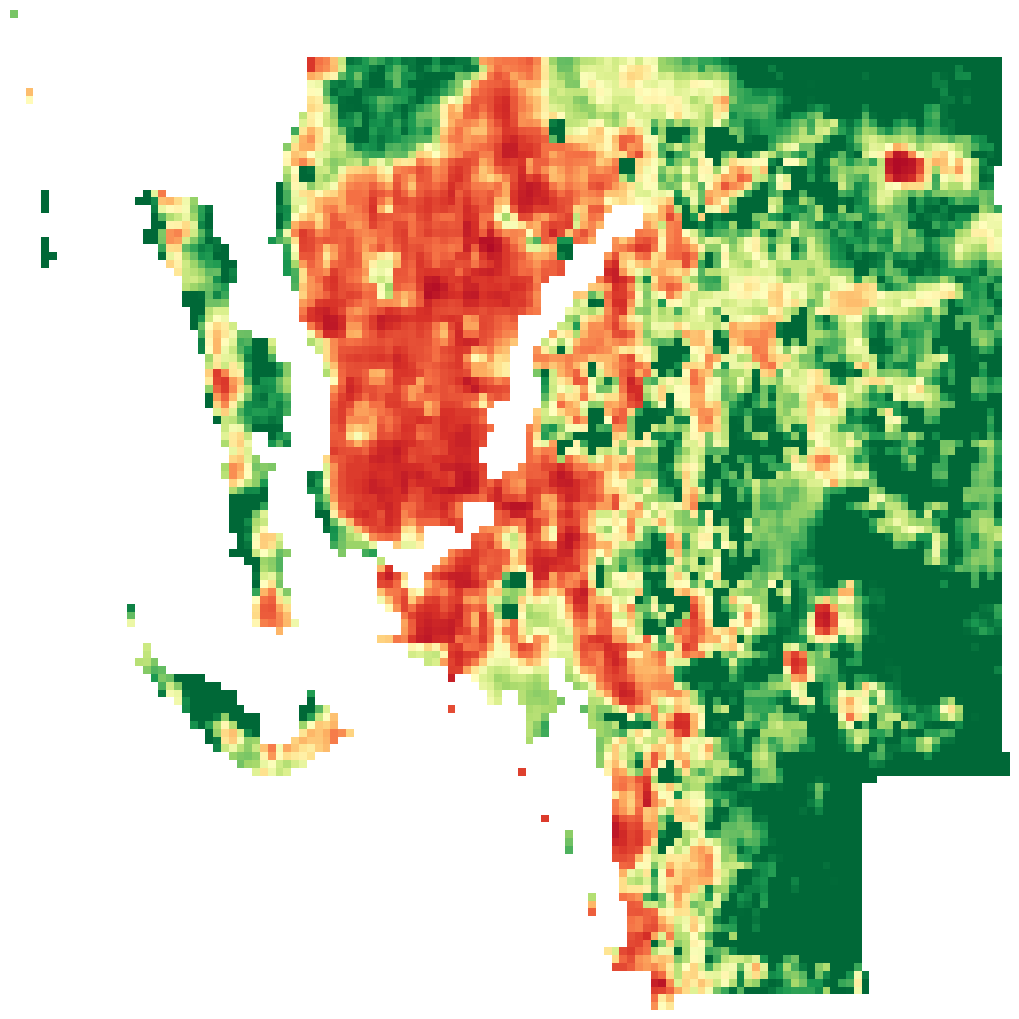}
            \end{minipage}
            \hfill
            \begin{minipage}{0.135\linewidth}
                \includegraphics[width=\linewidth]{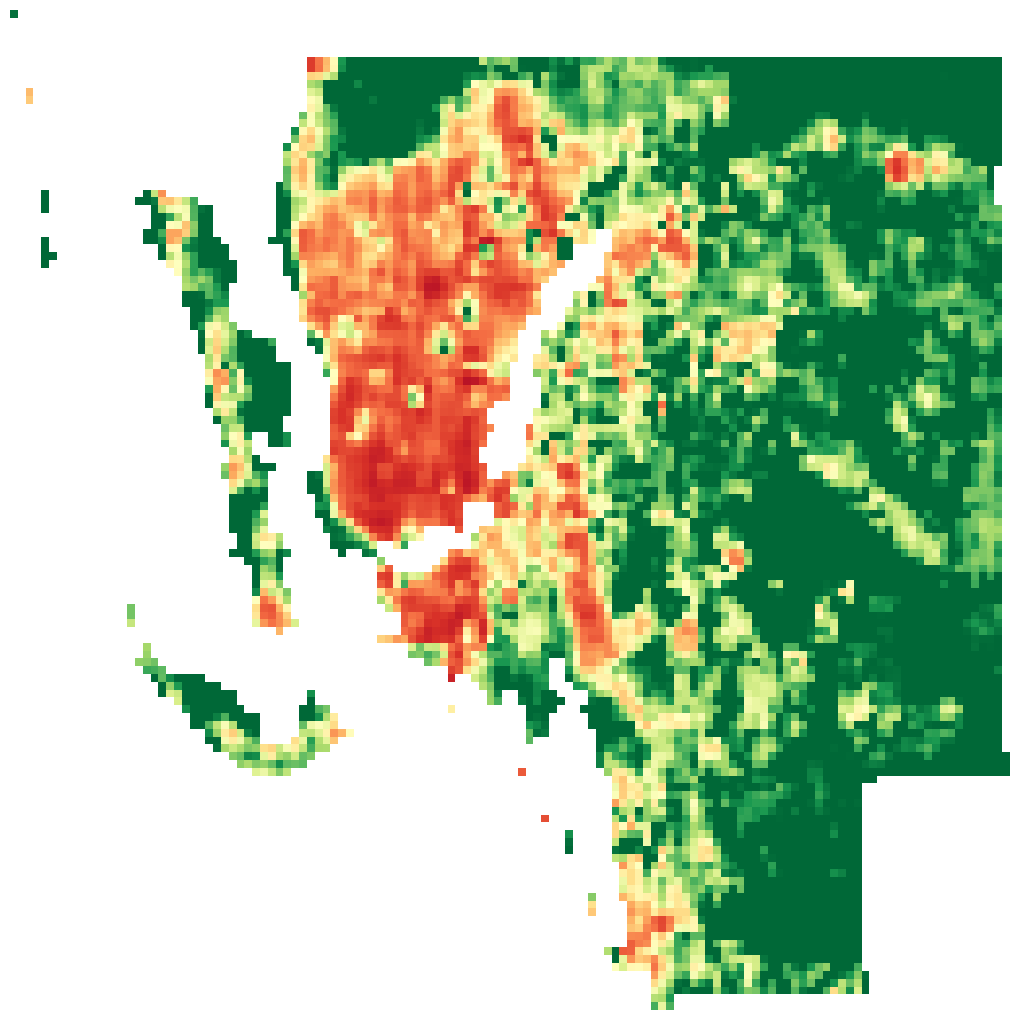}
            \end{minipage}
        
            \vspace{0.1cm}
        
            \begin{minipage}{0.135\linewidth}
                \includegraphics[width=\linewidth]{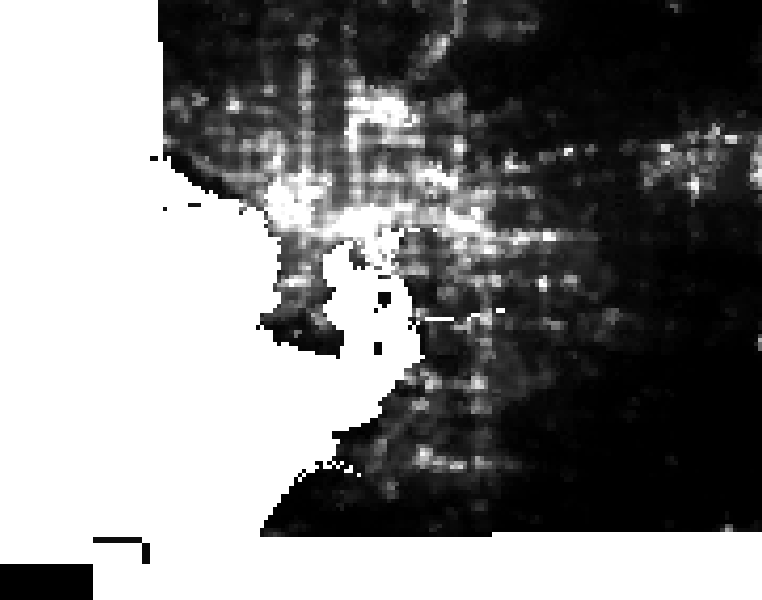}
            \end{minipage}
            \hfill
            \begin{minipage}{0.135\linewidth}
                \includegraphics[width=\linewidth]{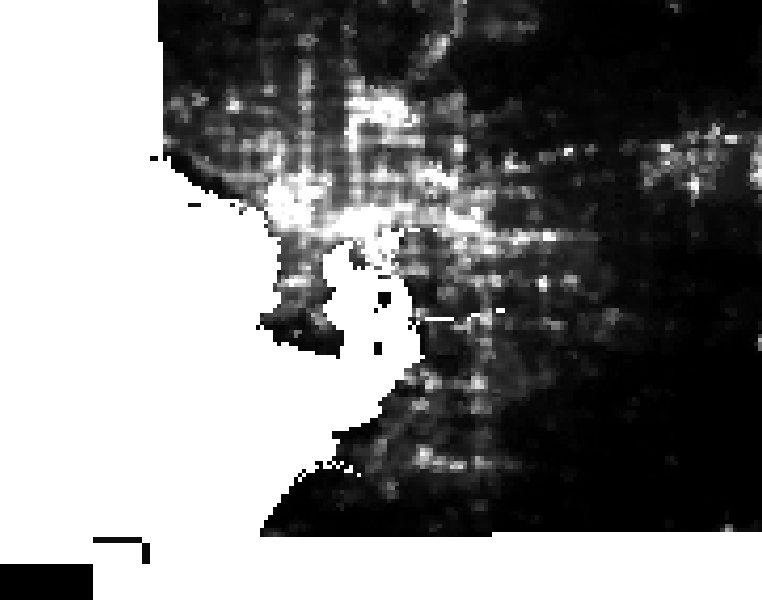}
            \end{minipage}
            \hfill
            \begin{minipage}{0.135\linewidth}
                \includegraphics[width=\linewidth]{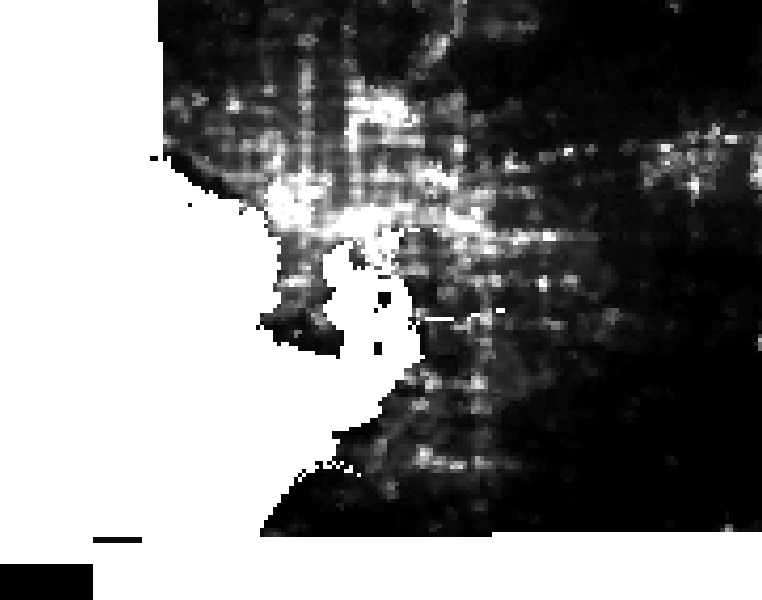}
            \end{minipage}
            \hfill
            \begin{minipage}{0.135\linewidth}
                \includegraphics[width=\linewidth]{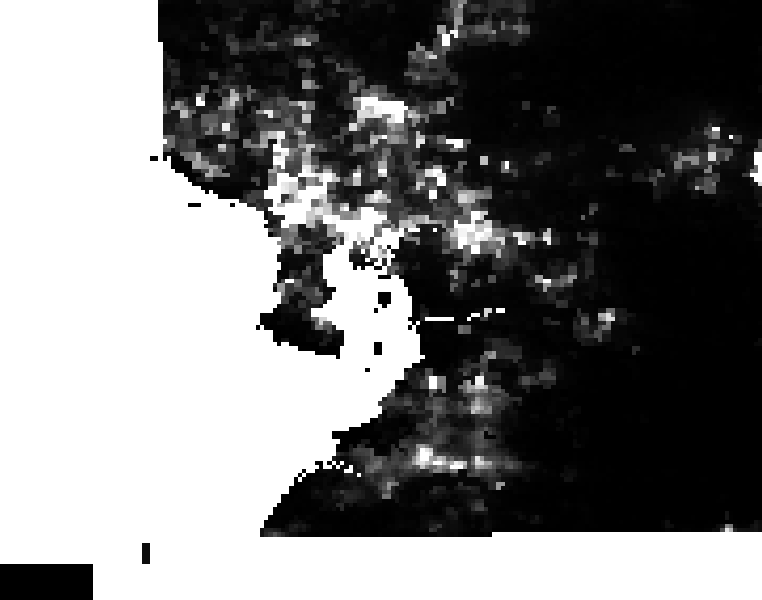}
            \end{minipage}
            \hfill
            \begin{minipage}{0.135\linewidth}
                \includegraphics[width=\linewidth]{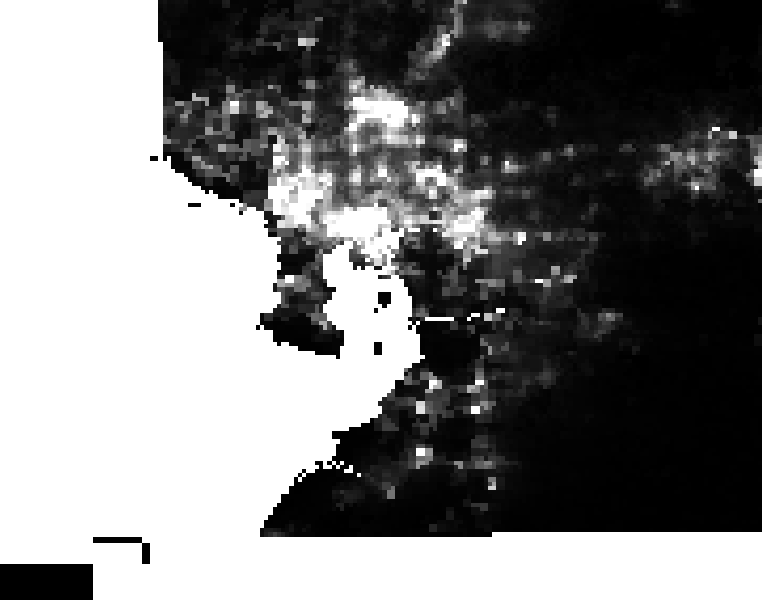}
            \end{minipage}
            \hfill
            \begin{minipage}{0.135\linewidth}
                \includegraphics[width=\linewidth]{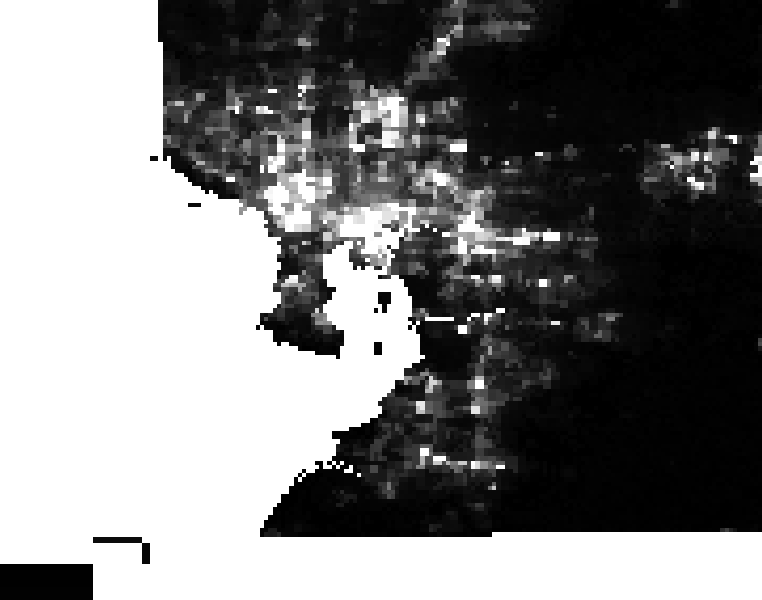}
            \end{minipage}
            \hfill
            \begin{minipage}{0.135\linewidth}
                \includegraphics[width=\linewidth]{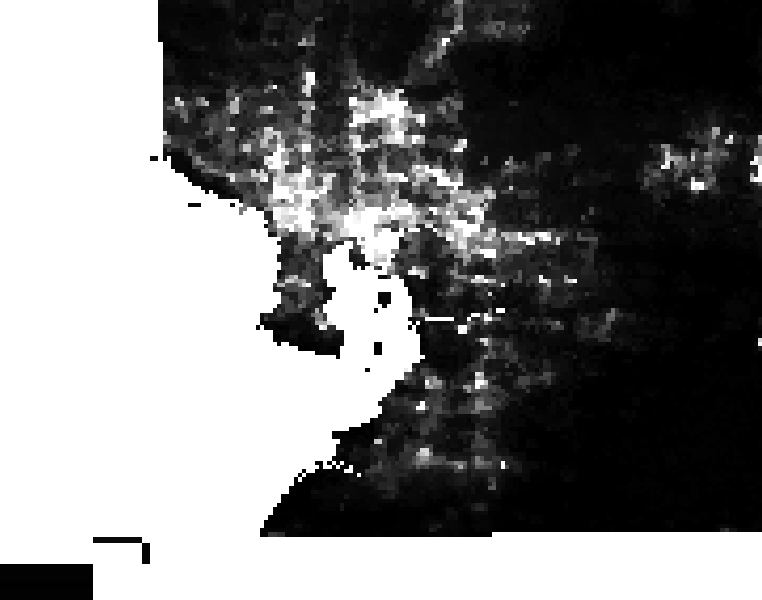}
            \end{minipage}
        
            \vspace{0.1cm}
        
            \begin{minipage}{0.135\linewidth}
                \includegraphics[width=\linewidth]{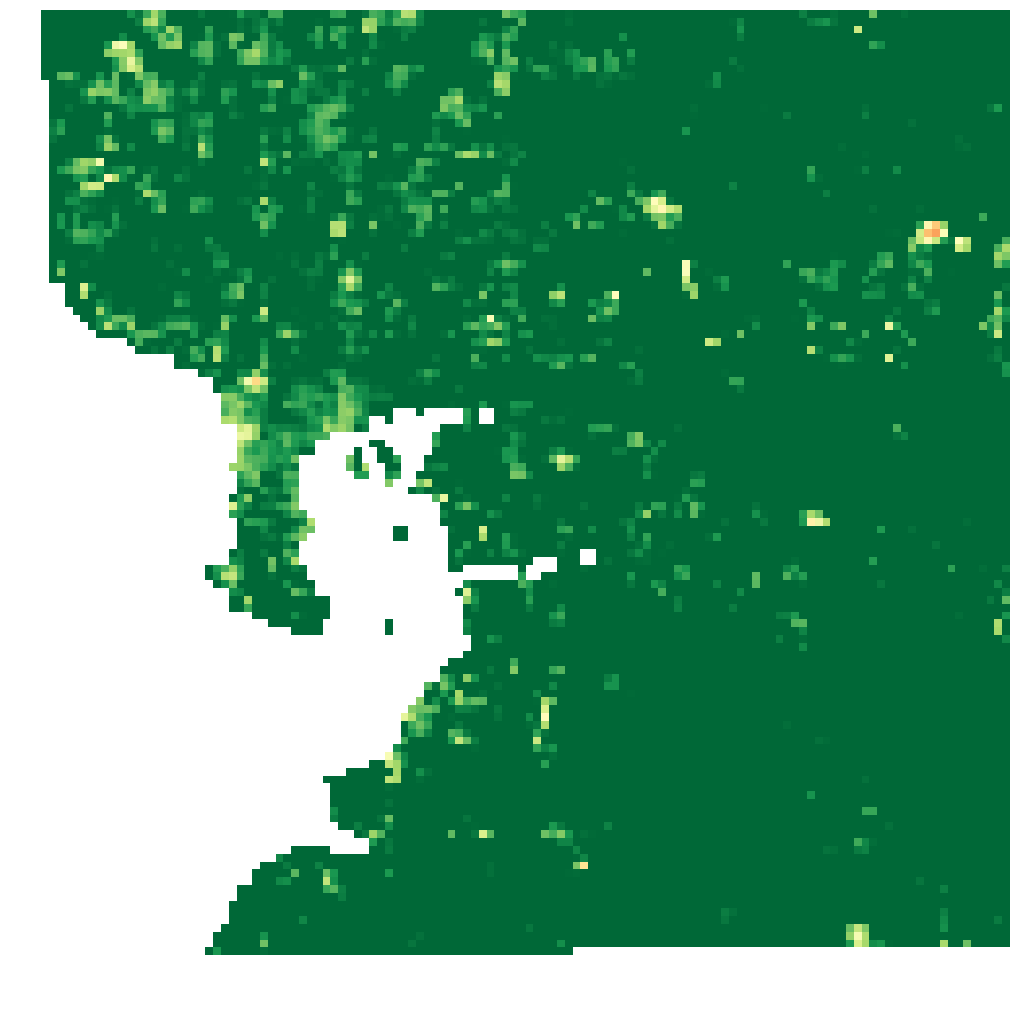}
            \end{minipage}
            \hfill
            \begin{minipage}{0.135\linewidth}
                \includegraphics[width=\linewidth]{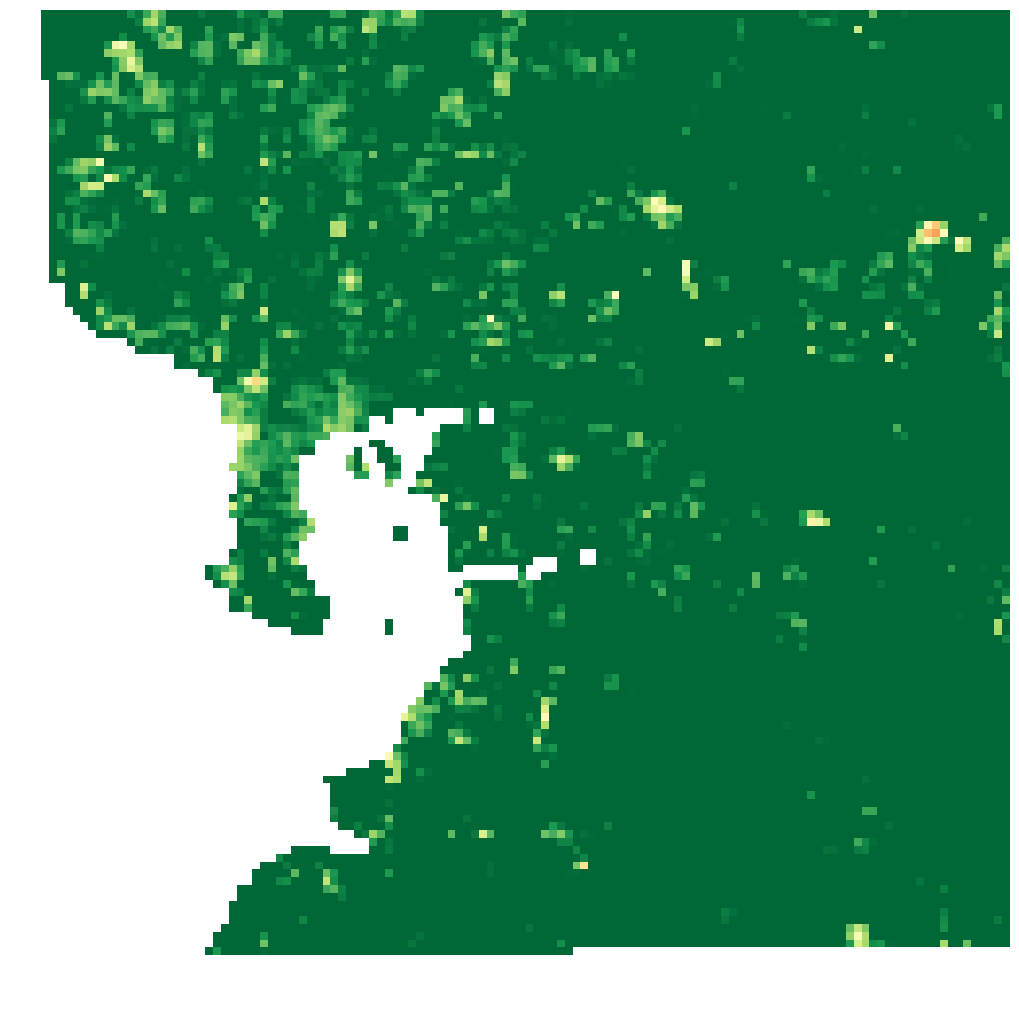}
            \end{minipage}
            \hfill
            \begin{minipage}{0.135\linewidth}
                \includegraphics[width=\linewidth]{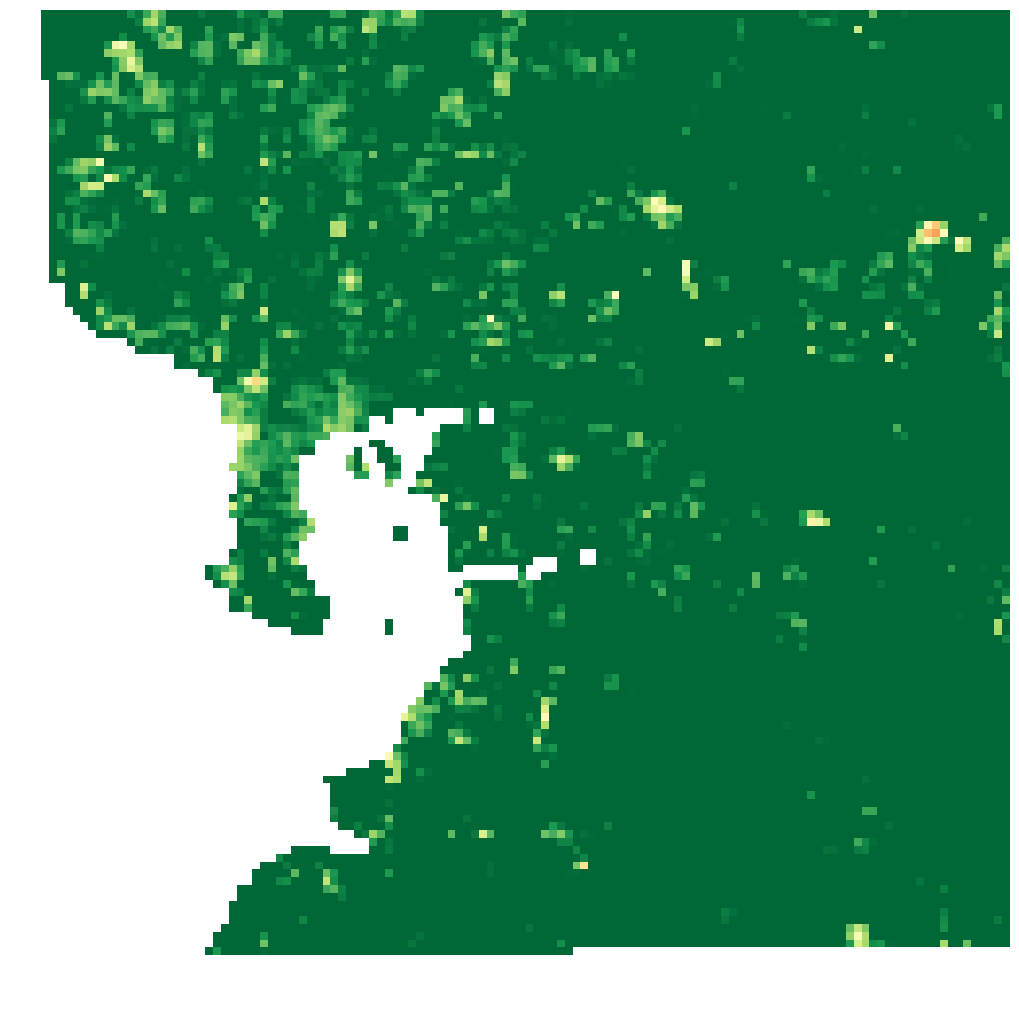}
            \end{minipage}
            \hfill
            \begin{minipage}{0.135\linewidth}
                \includegraphics[width=\linewidth]{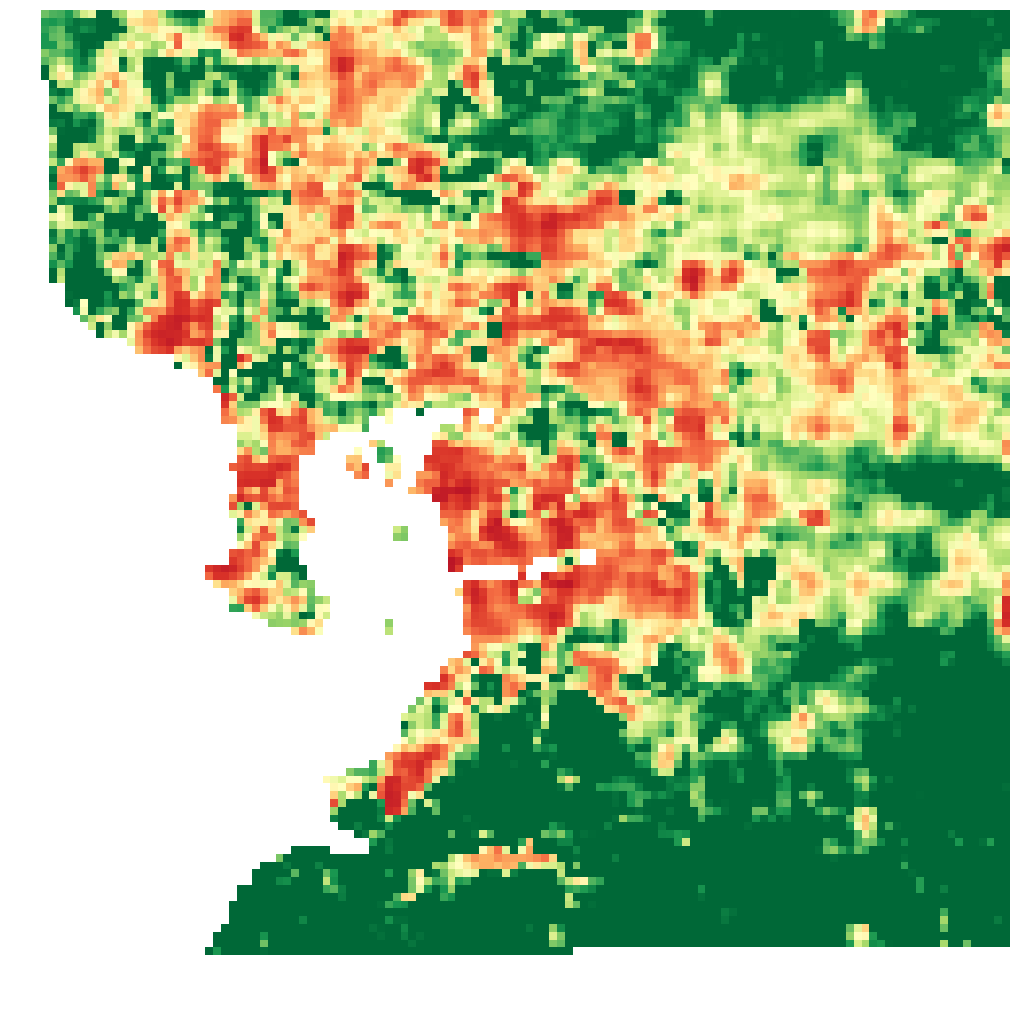}
            \end{minipage}
            \hfill
            \begin{minipage}{0.135\linewidth}
                \includegraphics[width=\linewidth]{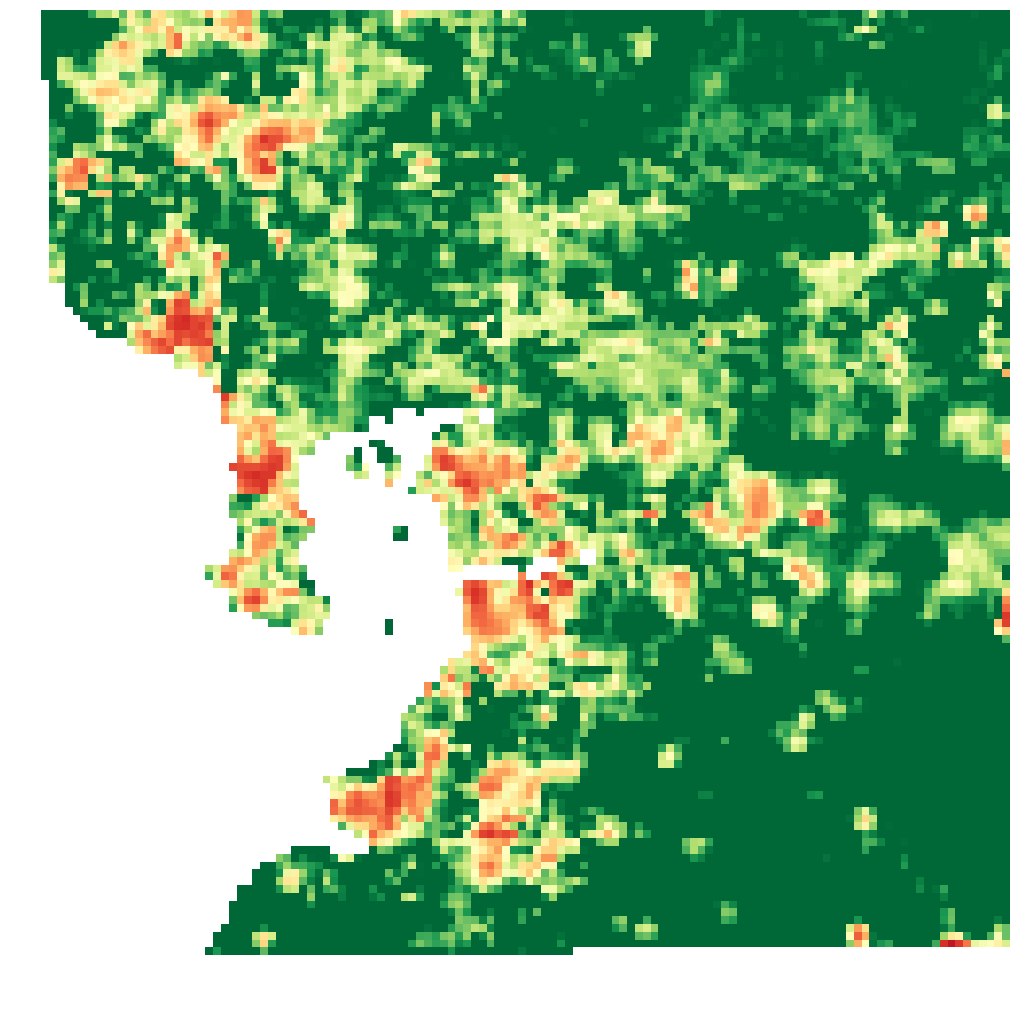}
            \end{minipage}
            \hfill
            \begin{minipage}{0.135\linewidth}
                \includegraphics[width=\linewidth]{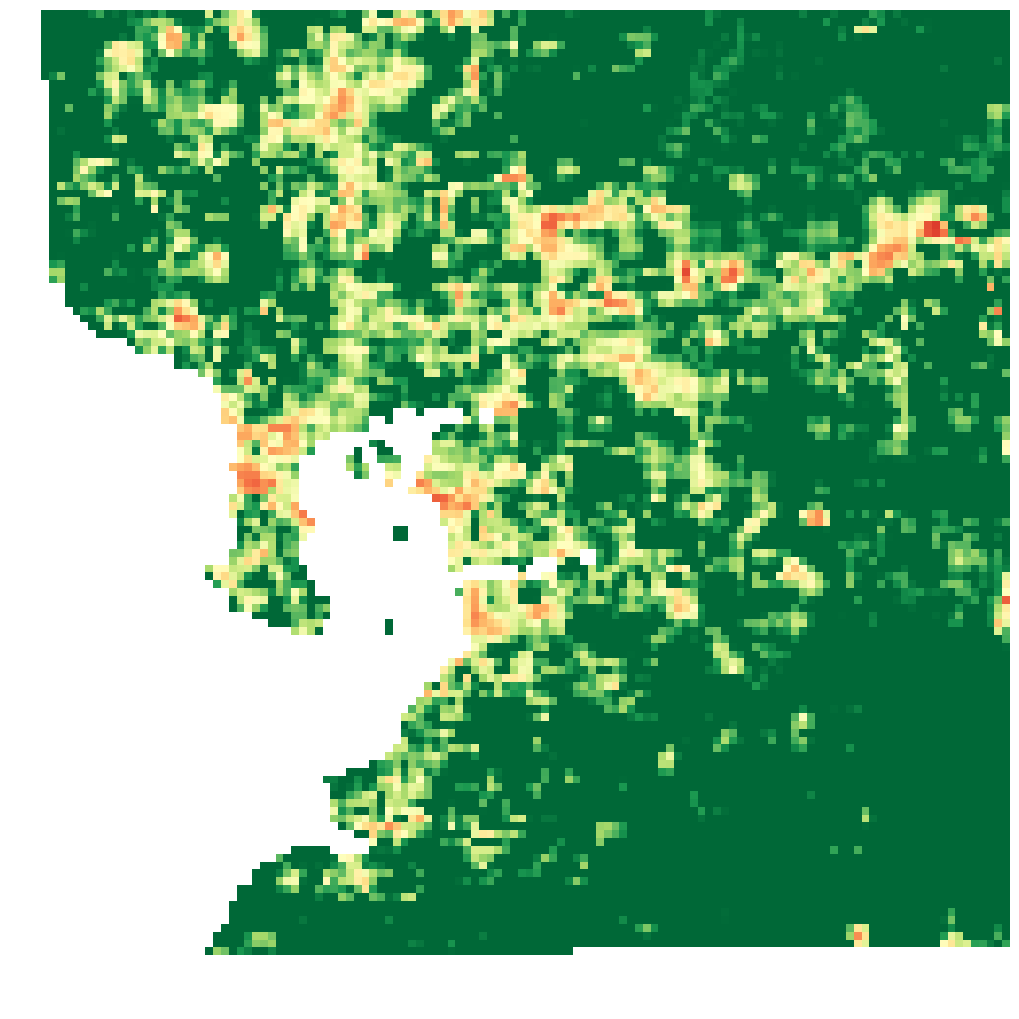}
            \end{minipage}
            \hfill
            \begin{minipage}{0.135\linewidth}
                \includegraphics[width=\linewidth]{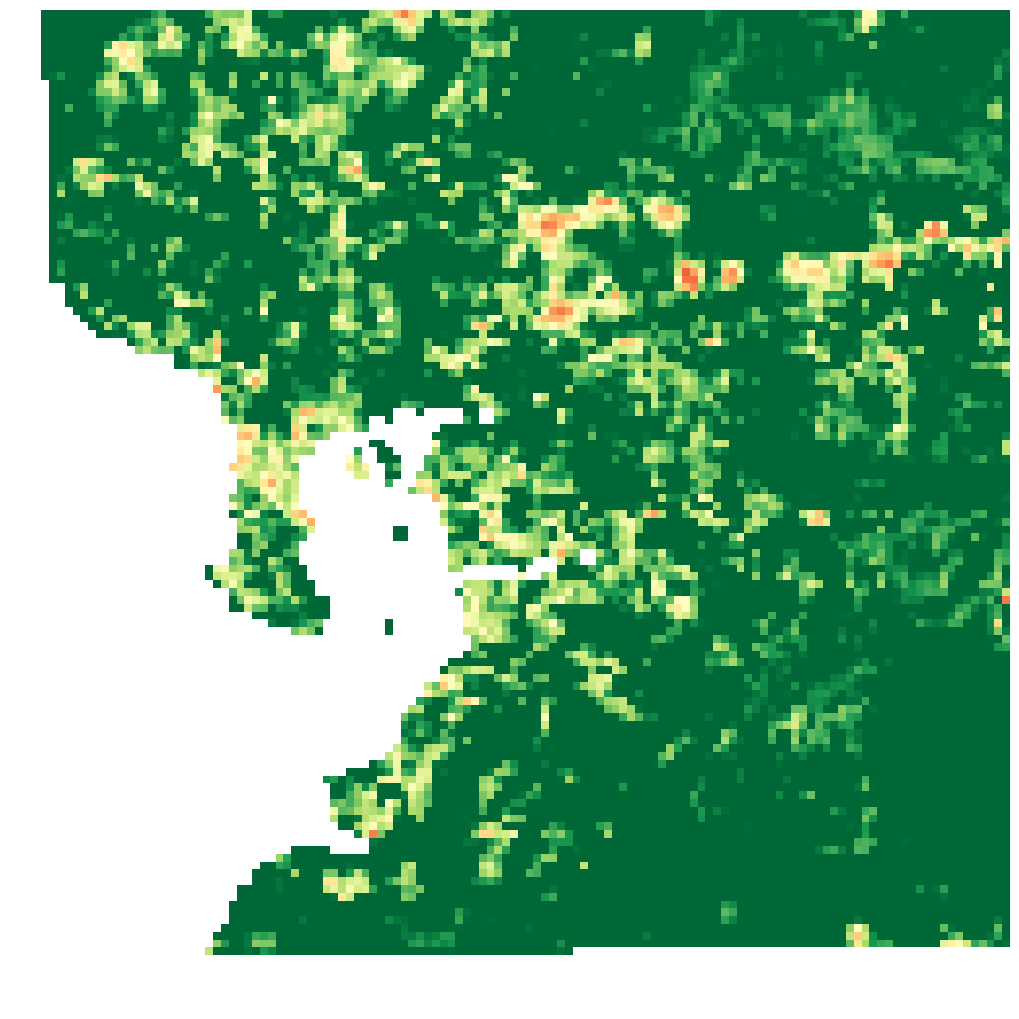}
            \end{minipage}
            \caption{Raw NTL images from the Black Marble dataset and the corresponding derived outage maps. 
            Rows 1 and 3 show the NTL images for Lee County, FL (during Hurricane Ian, September 26 – October 2, 2022) and Hillsborough County, FL (during Hurricane Milton, October 8 – 14, 2024), respectively. 
            Rows 2 and 4 present the corresponding outage maps for the same counties and time range.}
            \label{fig:ntloutageimages}
        \end{figure}

    The Black Marble dataset provides daily, high-resolution NTL satellite image data from NASA’s VIIRS sensor, capturing anthropogenic and natural light emissions to support applications such as disaster monitoring, energy access, and urbanization analysis \cite{blackmarble}.
    For Florida power outage analysis, we incorporate NTL image data from the daily VNP46A2 product, covering the entire state of Florida from January 2012 to January 2025. 
    Specifically, we extract the \textit{Gap\_Filled\_DNB\_BRDF-Corrected\_NTL} dataset from VNP46A2. 
    The raw data files are downloaded using the BlackMarblePy \cite{blackmarblepy} Python package and loaded as xarray DataArrays \cite{hoyer2017xarray}. 
    To further segment the data, we apply a bounding box for each Florida county to the NTL images, resulting in individual NTL images for each of Florida's 67 counties each night. 
    The image DataArrays for each county for each night are then saved in separate Pickle files. 
    Further metadata, such as the height and width of the resulting image and the date, are also extracted. 
    For generating the outage maps from NTL images, we use Aparcedo et al.'s \cite{aparcedo2024multimodalpoweroutageprediction} VST-GNN model ground-truth visualizations. 
    Outage maps are generated by comparing the daily radiance signal with the average radiance of the last three months for a given location \cite{aparcedo2024multimodalpoweroutageprediction,cole_2017,cui_2023,kalb_2018}.
    For each hurricane event analysis, we examine outage maps from one month before to one month after the hurricane's landfall.
    Figure \ref{fig:ntloutageimages} shows the raw NTL images and the corresponding derived outage maps for Hurricane Ian and Hurricane Milton, each spanning a one-week period. 
    Combining the two datasets, we see that the hurricane's landfall caused widespread power outages, followed by a gradual recovery over the subsequent days.

        \begin{table}[!tb]
              \centering
              \scriptsize
              \setlength{\tabcolsep}{4pt}
              \caption{EAGLE-I county-level outage records for Lee County, FL on 28 September 2022, between 12:00 and 16:00 UTC.}
              \begin{tabular}{rllrl}
                \toprule
                \textbf{fips\_code}
                  & \textbf{county}
                  & \textbf{state}
                  & \textbf{customers\_out}
                  & \textbf{run\_start\_time} \\
                \midrule
                12071 & Lee & Florida &  11\,244  & 2022-09-28 12:00:00 \\
                12071 & Lee & Florida &  13\,728  & 2022-09-28 12:15:00 \\
                12071 & Lee & Florida &  15\,798  & 2022-09-28 12:30:00 \\
                12071 & Lee & Florida &  23\,984  & 2022-09-28 12:45:00 \\
                12071 & Lee & Florida &  31\,307  & 2022-09-28 13:00:00 \\
                12071 & Lee & Florida &  30\,818  & 2022-09-28 13:15:00 \\
                12071 & Lee & Florida &  34\,751  & 2022-09-28 13:30:00 \\
                12071 & Lee & Florida &  34\,546  & 2022-09-28 13:45:00 \\
                12071 & Lee & Florida &  36\,676  & 2022-09-28 14:00:00 \\
                12071 & Lee & Florida &  38\,947  & 2022-09-28 14:15:00 \\
                12071 & Lee & Florida &  42\,785  & 2022-09-28 14:30:00 \\
                12071 & Lee & Florida &  43\,634  & 2022-09-28 14:45:00 \\
                12071 & Lee & Florida &  58\,227  & 2022-09-28 15:00:00 \\
                12071 & Lee & Florida &  72\,761  & 2022-09-28 15:15:00 \\
                12071 & Lee & Florida &  89\,363  & 2022-09-28 15:30:00 \\
                12071 & Lee & Florida & 103\,485  & 2022-09-28 15:45:00 \\
                12071 & Lee & Florida & 120\,112  & 2022-09-28 16:00:00 \\
                \bottomrule
              \end{tabular}
              \label{tab:lee_outages}
        \end{table}
    
    The EAGLE-I dataset is a publicly available U.S. Department of Energy resource that provides timeseries, county-level power outage data across the United States, aggregated from utility-reported information \cite{brelsford2024dataset,eaglei}.
    For our analysis, we filter the county-level outage data to include only Florida counties, covering the period from 2014 to 2024 at 15-minute intervals.
    We query the dataset to extract information on date and time ($run\_start\_time$), FIPS code ($fips\_code$), county and state name ($county$, $state$), and the number of customers without power ($customers\_out$).
    The given date and time ($run\_start\_time$) is converted to the RDF-standard $xsd:dateTime$\footnote[2]{YYYY-MM-DDTHH:MM:SSZ} format.
    Table \ref{tab:lee_outages} shows a snippet of the EAGLE-I \cite{brelsford2024dataset,eaglei} outage records for Lee County on 28 September 2022, during the landfall of Hurricane Ian \cite{h_ian}.
    We see a dramatic increase in the number of customers without power following the hurricane's landfall.

    To FAIRify \cite{jacobsenFAIRPrinciplesInterpretations2020,wilkinson_2016} the data, we create globally unique IRIs for all instances in the graph.
    For instance, \textit{ex:ntlimage.12001.2023-08-28} denotes a nighttime light image observation for Alachua County (FIPS code 12001) on August 28, 2023, while \textit{ex:outagerecord.12001.2023-08-28T00-00-00Z} represents an EAGLE-I outage record at midnight on the same date.
    This systematic approach guarantees that each instance is uniquely defined with respect to its spatial and temporal attributes.
    To ensure that the data will be interoperable and accessible, we developed a set of Resource Description Framework (RDF) \cite{w3cRDF12Schema2024} / Web Ontology Language (OWL) \cite{w3cOWL2Web2012} based vocabularies that connects established schema from, for example, GEOSatDB \cite{lin_2024}, DBpedia \cite{dbpedia}, and Ontology for Media Resources \cite{ma_ont}. 
    The data is made accessible through Turtle dump files, which are available on our OSF repository at \url{https://doi.org/10.17605/OSF.IO/QVD8B}. 
    The ontology and source code used to generate the knowledge graph is made publicly available through a GitHub repository at \url{https://purl.org/geooutagekg}.

    \subsection{GeoOutageOnto: Ontology Development} \label{subsec3.3}
    
        \begin{figure}[t]
            \centering
            \includegraphics[scale=0.4]{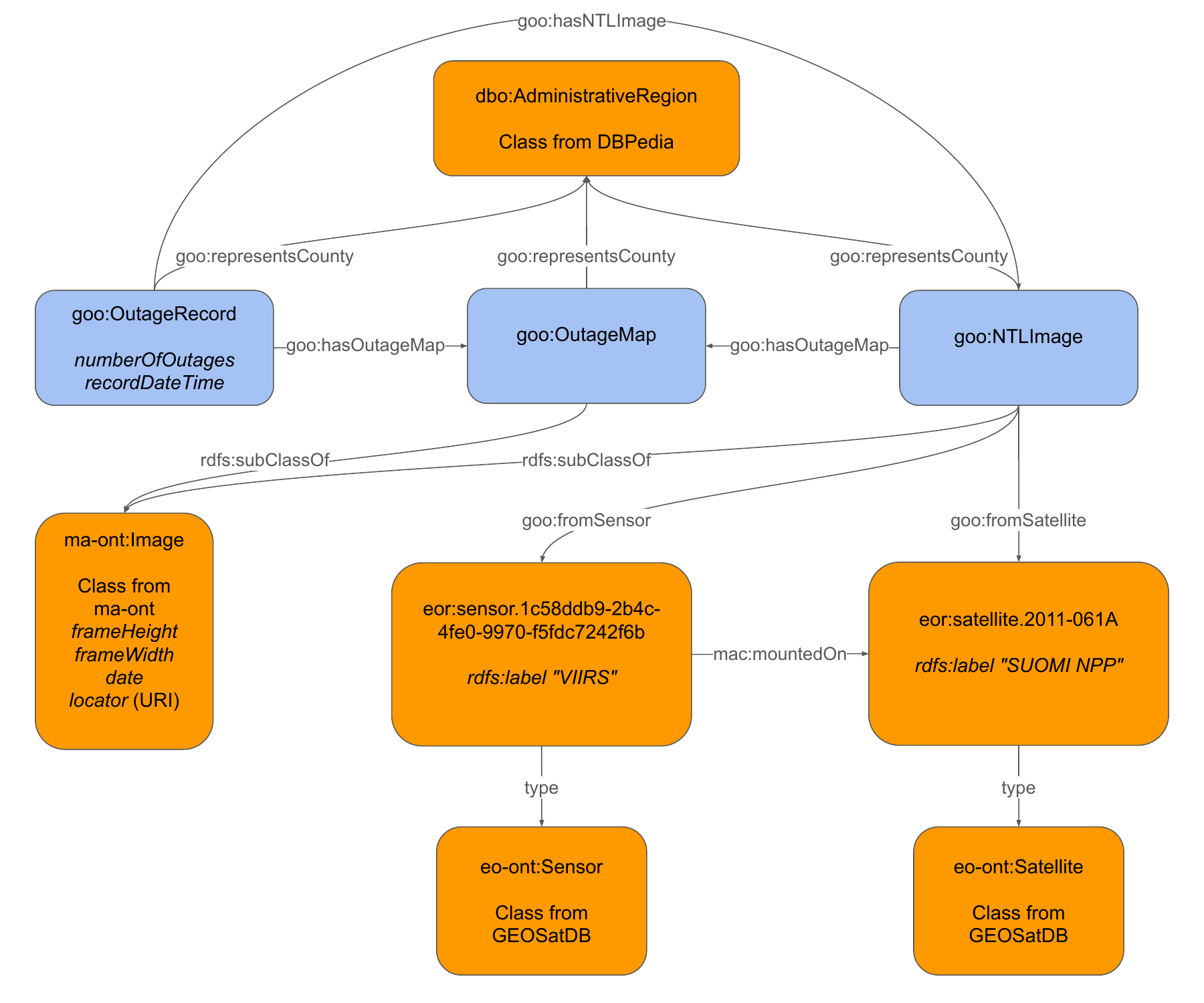}
            \caption{Visualization of GeoOutageOnto, with classes and relations displayed. Novel classes are shown in blue, while reused classes from other ontologies are shown in orange. Class or instance attributes are italicized. GeoOutageOnto reuses several ontologies, most notably DBpedia, GeoSatDB, and Ontology for Media Resources.}
            \label{fig:ontoutline}
        \end{figure}
        
        GeoOutageOnto is designed to integrate heterogeneous data sources into a unified semantic representation, enabling more effective geospatiotemporal power outage analysis. 
        Drawing on domain expertise and familiarity with existing datasets—such as satellite image data, time series data, and geospatiotemporal outage maps—we identified domain-relevant concepts, properties, and relationships. 
        Each data modality is modeled as a distinct subontology class to support modular integration. 
        We also reuse schema elements from well-established ontologies to promote semantic interoperability and alignment with existing linked data standards.
        
        $OutageRecord$. This subontology covers the representation of timeseries data; the number of customers with power outages recorded in the EAGLE-I dataset \cite{brelsford2024dataset,eaglei}, for example.

        $OutageMap$. This subontology covers the representation of geospatiotemporal outage severity maps.

        $NTLImage$. This subontology covers the representation of satellite image data, such as county-masked images from the Black Marble dataset \cite{blackmarble}. 
        
        Figure \ref{fig:ontoutline} provides an overview of the structure of GeoOutageOnto. The proposed new subontology classes are highlighted in blue and the reused terms are shown in orange. 
        To include county information, we link to county instances from DBpedia's \cite{dbpedia} $AdministrativeRegion$ class, which contains statistical metadata such as the region's name, population, and land/water area. 
        Each subontology that includes a geospatial component links to its corresponding county IRI via the predicate $representsCounty$. 
        Satellite image data contain rich information on the type of satellite and sensor. 
        For example, the Black Marble dataset contains information from the Suomi NPP satellite and VIIRS sensor. 
        To integrate the satellite image data, we link to the GEOSatDB \cite{lin_2024} knowledge graph.
        The $NTLImage$ subontology, containing links to Black Marble satellite image data, has a $fromSatellite$ predicate to link to GEOSatDB's Suomi NPP satellite instance, and a $fromSensor$ predicate to link to GEOSatDB's VIIRS sensor instance. 
        The sensor is further connected to the satellite in GEOSatDB itself through the $mountedOn$ predicate. 
        To classify the $NTLImage$ and $OutageMap$ instances as an image instance, we define them as subclasses of the Image class from the Ontology for Media Resources \cite{ma_ont}. 
        This Image class is in turn a subclass of MediaResource, which contains metadata such as the image's height and width, date/time, and IRI. 
        Linking diverse data sources within GeoOutageKG enables comprehensive outage analysis and supports downstream applications, such as grid resilience assessment, energy access disparity evaluation, and community energy vulnerability analysis.

\section{GeoOutageKG Statistics} 
    \label{sec4}  
    
    At the time of writing, GeoOutageKG contains 3 primary datasets with over 10.9 million instances and over 88.9 million RDF triples (or semantic statements). 
    Table \ref{tab:num_instances} shows the statistics of the $Outage Record$, $NTL Image$ and $Outage Map$ instances. 

    \begin{table}[!tb]
          \centering
          \scriptsize
          \setlength{\tabcolsep}{4pt}
          \caption{Number of instances and semantic statements (RDF triples) per class in GeoOutageKG.}
          \begin{tabular}{rllrl}
            \toprule
            \textbf{Class}
              & \textbf{Instances}
              & \textbf{Semantic Statements} \\
            \midrule
            $OutageRecord$ & 10\,635\,995 & 85\,679\,249 &  \\
            $NTLImage$ & 313\,702 & 3\,152\,564   &  \\
            $OutageMap$ & 15\,544 & 139\,896   &  \\
            \midrule
            \textbf{Total} & 10\,965\,241 & 88\,971\,709 \\
            \bottomrule
          \end{tabular}
          \label{tab:num_instances}
    \end{table}

    \begin{table}[!tb]
          \centering
          \scriptsize
          \setlength{\tabcolsep}{4pt}
          \caption{Number of instances of the $NTLImage$ class per year. Note that some years have missing days due to gaps in dataset.}
          \begin{tabular}{r r r}
            \toprule
            \textbf{Year}
              & \textbf{Total Instances}
              & \textbf{Mean Instances per County} \\
            \midrule
            2012 & 22\,542  & 336.45  \\
            2013 & 24\,455  & 365.00  \\
            2014 & 24\,254  & 362.00  \\
            2015 & 24\,418  & 364.45  \\
            2016 & 24\,403  & 364.22  \\
            2017 & 24\,284  & 362.45  \\
            2018 & 24\,455  & 365.00  \\
            2019 & 24\,321  & 363.00  \\
            2020 & 24\,485  & 365.45  \\
            2021 & 24\,284  & 362.45  \\
            2022 & 23\,316  & 348.00  \\
            2023 & 24\,134  & 360.21  \\
            2024 & 22\,311  & 333.00  \\
            2025 & 2\,040 & 30.45 \\
            \midrule
            \textbf{Total} & 313\,702 & 4\,682.12 \\
            \bottomrule
          \end{tabular}
          \label{tab:num_ntl_instances}
    \end{table}

    \subsection{Satellite Image Statistics} 
    \label{subsec4.1}
    
    The current satellite image data integrated into GeoOutageKG contains NTL Images from the VNP46A2 product collected through NASA’s VIIRS sensor, which is mounted on the Suomi NPP satellite. 
    The full VNP46A2 dataset exceeds 252GB in size. 
    After spatially filtering the data to Florida for our case study, the dataset is reduced to about 5.1 GB. 
    GeoOutageKG contains 313,702 individual $NTLImage$ entries, each representing a nightly observation for a specific county in Florida from January 2012 to January 2025. 
    The distribution of $NTLImage$ entries by year and by county per year is summarized in Table \ref{tab:num_ntl_instances}.

    \subsection{Outage Map Statistics} 
    \label{subsec4.2}
    
    Currently, the $OutageMap$ class integrated into GeoOutageKG contains 15,544 outage maps. 
    These maps are derived from nighttime light satellite images associated with the five most recent major hurricane events in Florida. 
    The $OutageMap$ class contributes a total of 139,896 triples.
    This corresponds to approximately 3,886 maps per hurricane, spanning all 67 counties in Florida.
    On average, there are 58 outage maps per county for each hurricane.\footnote[3]{As Hurricanes Helene and Milton occurred within two weeks of each other, their data is grouped as one hurricane.}

    \begin{table}[!tb]
          \centering
          \scriptsize
          \setlength{\tabcolsep}{4pt}
          \caption{Number of instances of the $OutageRecord$ class per county.}
          \begin{tabular}{lr lr lr}
            \toprule
            County       & Instances   & County       & Instances   & County       & Instances   \\
            \midrule
            Alachua      & 188\,957     & Baker        &  54\,324     & Bay          & 185\,867     \\
            Bradford     &  50\,117     & Brevard      & 311\,776     & Broward      & 349\,398     \\
            Calhoun      &  32\,278     & Charlotte    & 239\,224     & Citrus       & 171\,231     \\
            Clay         & 108\,418     & Collier      & 283\,643     & Columbia     & 122\,859     \\
            DeSoto       & 109\,352     & Dixie        &  54\,731     & Duval        & 216\,839     \\
            Escambia     & 234\,205     & Flagler      & 136\,194     & Franklin     &  59\,050     \\
            Gadsden      &  93\,975     & Gilchrist    &  64\,510     & Glades       &  37\,513     \\
            Gulf         &  75\,771     & Hamilton     &  54\,691     & Hardee       &  60\,175     \\
            Hendry       &  97\,057     & Hernando     & 153\,299     & Highlands    & 131\,633     \\
            Hillsborough & 300\,312     & Holmes       &  45\,735     & Indian River & 226\,452     \\
            Jackson      &  59\,386     & Jefferson    &  59\,153     & Lafayette    &  50\,329     \\
            Lake         & 202\,407     & Lee          & 298\,123     & Leon         & 182\,603     \\
            Levy         & 103\,402     & Liberty      &  47\,138     & Madison      &  41\,506     \\
            Manatee      & 270\,944     & Marion       & 233\,530     & Martin       & 223\,805     \\
            Miami-Dade   & 350\,232     & Monroe       &  55\,031     & Nassau       & 110\,409     \\
            Okaloosa     & 169\,877     & Okeechobee   & 109\,034     & Orange       & 302\,805     \\
            Osceola      & 152\,107     & Palm Beach   & 341\,342     & Pasco        & 254\,663     \\
            Pinellas     & 315\,675     & Polk         & 263\,189     & Putnam       & 139\,434     \\
            Santa Rosa   & 169\,800     & Sarasota     & 294\,157     & Seminole     & 258\,617     \\
            St.\ Johns   & 195\,867     & St.\ Lucie   & 250\,050     & Sumter       & 117\,500     \\
            Suwannee     & 127\,427     & Taylor       &  47\,523     & Union        &  27\,577     \\
            Volusia      & 292\,018     & Wakulla      &  85\,629     & Walton       & 106\,500     \\
            Washington   &  77\,620     &              &              &              &              \\
            \bottomrule
          \end{tabular}
          \label{tab:num_record_instances}
    \end{table}

    \subsection{Outage Record Statistics} 
    \label{subsec4.3}
    
    GeoOutageKG currently contains a total of 10,635,995 $OutageRecord$ instances.
    This timeseries dataset captures county-level power outage information filtered for the state of Florida from 2014 to 2024. 
    Each year, the dataset includes approximately 900,000 and 1.2 million individual $OutageRecord$ entries.
    The complete list of Florida counties and their respective counts of $OutageRecord$ instances is provided in Table \ref{tab:num_record_instances}.

    \begin{figure}[!tb]
      \centering
      \subfloat[]{%
        \includegraphics[width=0.48\textwidth]{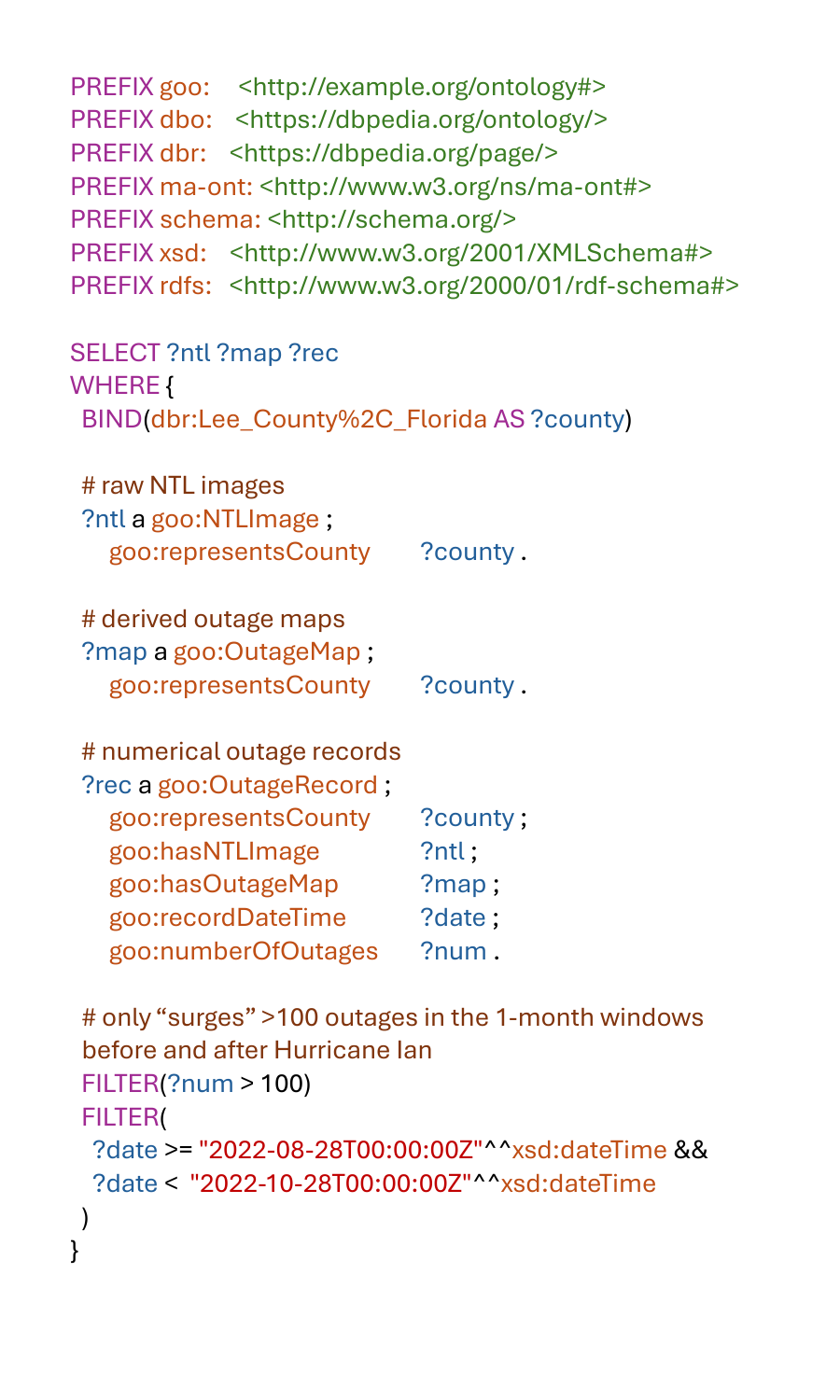}%
        \label{fig:sparqlquery}%
      }%
      \hfill
      \subfloat[]{%
        \includegraphics[width=0.48\textwidth]{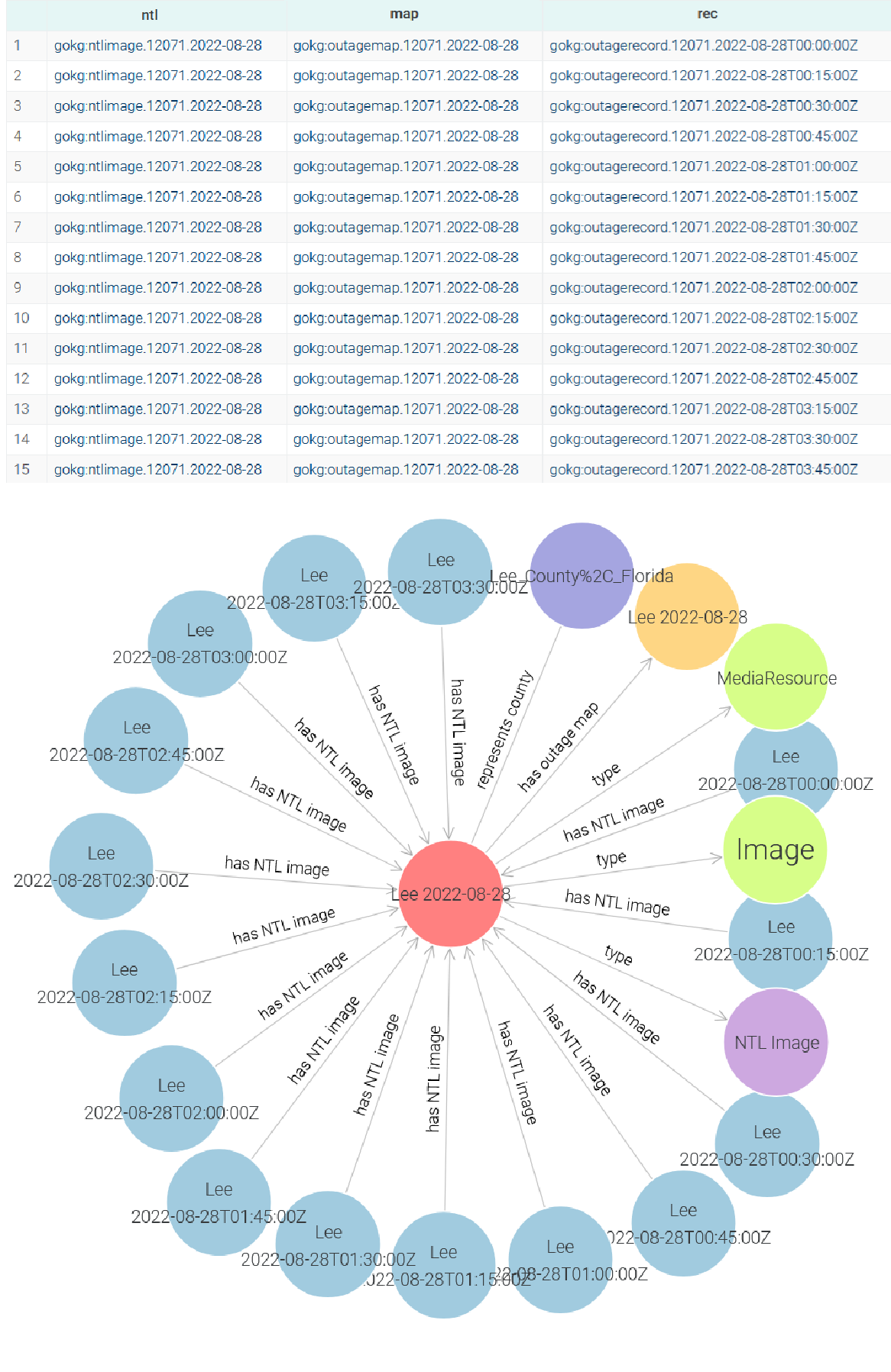}%
        \label{fig:sparqlresult}%
      }%
      \caption{(a) SPARQL query for retrieving outage records, NTL Images, and outage maps around Hurricane Ian in 2022. (b) Raw result and visualization of the query: counties with >100 outages. The first 15 results are displayed.}
      \label{fig:combined-sparql}
    \end{figure}

\section{Applications and Use Cases} 
    \label{sec5}
    
    In this section, we demonstrate a range of use cases enabled by GeoOutageKG, highlighting its utility for spatiotemporal outage analysis, energy access disparity assessment, and grid operation decision support. 

    \subsection{Power Outage Severity Evaluation} 
    \label{subsec5.1}
    
    One core use case of GeoOutageKG is the evaluation of power outage severity following extreme weather events. 
    Accurate severity assessment is essential for grid risk evaluation, emergency response, and disaster recovery planning. 
    Traditional outage data offer fine-grained temporal resolution, but are typically aggregated at the county level, limiting their spatial expressiveness. 
    Conversely, satellite image data provides detailed spatial information capable of revealing sub-county outage patterns but is limited to daily observations. 
    GeoOutageKG semantically integrates these complementary data modalities—high-frequency time series outage reports and high-resolution spatial imagery—into a unified knowledge graph. 
    This integration enables the joint analysis of outage extent, intensity, and duration with explicit spatial and temporal grounding. 
    By linking individual $OutageRecord$ instances to their corresponding $NTLImage$ and $OutageMap$, analysts can derive a more granular and holistic view of outage severity across both space and time. 
    This structured representation also supports downstream tasks such as reasoning over affected infrastructure, assessing recovery trajectories, and informing resilience strategies.
    Figure \ref{fig:combined-sparql} shows an example of a SPARQL query of GeoOutageKG, retrieving data for Lee County during Hurricane Ian. 

    \subsection{Energy Access Disparity and Vulnerability Analysis} 
    \label{subsec5.2}
    
    GeoOutageKG enables spatiotemporal analysis of historical power outage patterns, facilitating insights into long-term disparities in energy access and grid reliability. 
    By linking high-resolution outage data across multiple hurricane events and daily nighttime light (NTL) imagery, the knowledge graph reveals which communities experience more frequent or prolonged outages over time. 
    This integrated representation allows stakeholders to identify structurally vulnerable regions—such as rural or underserved communities—that may require targeted infrastructure investments or resilience planning. 
    Furthermore, the semantic alignment of outage records, geographic identifiers, and event metadata within GeoOutageKG supports comparative analysis across events and locations, helping to prioritize areas with persistent energy access challenges.

    \subsection{Grid Operation Decision Support} 
    \label{subsec5.3}
    
    GeoOutageKG serves as a semantic foundation for advanced decision support in energy distribution, enabling the extraction of localized, spatiotemporally conditioned statistical patterns to inform grid management and resilience planning.
    The knowledge graph allows users to derive context-specific "normal" ranges of power consumption or outage behavior.
    For example, regions that exhibit similar patterns of visual light intensity, population density, and administrative profiles may also demonstrate predictable ranges of outage duration or power consumption.
    Similarly, in commercial-dense areas, delayed effects of outages—such as sustained post-event power consumption spikes—can be inferred and modeled.
    These contextual rules, encoded within the ontology or inferred from linked data, can be used to validate model outputs, detect anomalies, or constrain predictive reasoning.
    For example, a predicted power consumption value that falls outside historically consistent bounds for a given spatiotemporal context can be flagged for expert review. 
    GeoOutageKG thus functions not only as a data repository but as a configurable semantic layer that supports explainable and domain-informed power grid decision making.

\section{Maintenance, Limitations, and Future Work} 
    \label{sec6}

    To maintain the longevity of GeoOutageKG, we plan for regular updates to existing data classes by adding new instances for recently published data. 
    $NTLImage$ instances, for example, temporally span from January 2012 to January 2025 as of current. 
    As the Black Marble \cite{blackmarble} dataset is regularly updated, however, frequent updates with new dates will help maintain data recency and expand data temporal range.
    Thus, we intend to add new $NTLImage$ data every 3 months.
    While the EAGLE-I \cite{brelsford2024dataset,eaglei} dataset is not updated as frequently, only annually, we also intend to update GeoOutageKG annually with the previous year's $OutageRecord$ data upon release.
    
    While GeoOutageKG currently supports meaningful use cases in outage detection and risk assessment, there remains significant potential for future expansion. 
    At present, the ontology models three primary data classes: $NTLImage$, $OutageRecord$, and $OutageMap$. These classes are sufficient for spatiotemporal analysis, but additional datasets could enhance both granularity and coverage.
    For example, the integration of meteorological satellite data, such as the NOAA GOES series \cite{goes}, would provide a valuable context for disaster-related outage patterns and would enable more robust modeling of extreme weather-related grid disruptions.
    Similarly, extending the ontology to include infrastructure connectivity, restoration prioritization rules, and hazard exposure models could further support advanced reasoning tasks and operational decision-making.
    We also plan to expand the knowledge graph by incorporating additional datasets, such as infrastructure asset data, distributed energy resources (DER) deployments, and sociodemographic indicators, to support broader use cases in energy justice and disaster resilience.
    The GeoOutageKG ontology and codebase are publicly available. 
    We welcome community contributions that help align GeoOutageKG with evolving domain ontologies and standards, particularly those in geospatial, energy, and disaster risk management domains.
    GeoOutageKG is designed in collaboration with and is the geospatial component of the MDS-Onto \cite{rajamohan_2025} ontology project. 
    Ontology interfacing and contribution is available through MDS-Onto API.

    Additionally, as of current, GeoOutageKG contains geospatiotemporal data within the state of Florida only. 
    GeoOutageKG's modularity, however, permits its spatial resolution to expand not only across the entire US, but any national or subnational administrative region across the world.
    While its current data corpus is large, the potential of rapidly expanding it by including county-level data from other U.S. states or national subdivisions is currently being explored.
    Moreover, more geospatial data is being incorporated directly into the ontology and knowledge graph themselves, as accessing the geospatial data for each county and image currently relies on accessing the external DBpedia knowledge graph \cite{dbpedia}. 
    Data such as the bounding box coordinates of each county has been extracted directly from the segmented NTL images, and further preprocessing such as the utilization of geohashing is also being explored. 
    One method being explored is directing the $representsCounty$ predicate to a new $County$ class.
    This $County$ class will then link to its geospatial coordinate and geohash data, as well as external resources such as its 
    DBpedia \cite{dbpedia} and Wikidata \cite{wikidata2025} entries.

    GeoOutageKG is also being explored for potential applications in Large Language Model (LLM) benchmarking and retrieval. 
    Methods such as OG-RAG \cite{sharma_2024} have enhanced LLM text generation by querying and matching domain-specific ontologies relevant to the prompt using Retrieval-Augmented Generation (RAG). 
    Such a method could also be inspiration for a RAG pipeline using GeoOutageKG, where users can query a LLM in a natural language format and receive outage data in return.
    
\section{Conclusion} 
    \label{sec7}

    In this paper, we introduced GeoOutageKG, a multimodal, ontology-derived knowledge graph designed to support geospatiotemporal analysis of power outages. 
    By integrating high-resolution nighttime light satellite image data, time-series outage records, and derived outage severity maps, GeoOutageKG enables richer semantic representation and reasoning across multiple data modalities and temporal scales. 
    The underlying ontology, GeoOutageOnto, provides a structured schema for linking outage events to geographic and temporal context, facilitating advanced use cases in disaster impact analysis, infrastructure resilience, and energy access disparity assessment.
    We highlighted how GeoOutageKG supports use cases such as outage severity evaluation, energy vulnerability analysis, and grid operation decision support.
    Looking ahead, we plan to expand GeoOutageKG by incorporating additional datasets, including meteorological data, infrastructure assets, and sociodemographic indicators. 
    We also aim to enhance its reasoning capabilities through the integration of localized statistical rules and predictive validation models. 
    As power grids face increasing stress from extreme weather and climate-related disruptions, we envision GeoOutageKG as a reusable and extensible semantic infrastructure to support equitable, data-driven energy system planning and disaster mitigation.

\paragraph{\normalfont\bfseries Resource Availability Statement.}
    
    GeoOutageKG and the data collected for creating it are licensed under the Creative Commons Attribution 4.0 International (CC BY 4.0) license, available to freely use and reuse for all noncommercial purposes. 
    The source code used to generate the knowledge graph, as well as the documentation for GeoOutageKG, is available from GitHub at \url{https://purl.org/geooutagekg}.
    The source code is available under the MIT License.
    Additionally, all image files, Turtle dump files, and other metadata can be found on our OSF repository at \url{https://doi.org/10.17605/OSF.IO/QVD8B}.
    The documentation for GeoOutageOnto can be found at \url{https://purl.org/geooutageonto}.

\paragraph{\normalfont\bfseries Acknowledgements.}
    
    This material is based upon research in the Materials Data Science for Stockpile Stewardship Center of Excellence (MDS3-COE), and supported by the Department of Energy's National Nuclear Security Administration under Award Number(s) DE-NA0004104. 
    This material is based upon work supported by the U.S. National Science Foundation, Office of Advanced Cyberinfrastructure (OAC), Major Research Instrumentation, under Award Number 2117439. 
    This work made use of the High Performance Computing Resource in the Core Facility for Advanced Research Computing at Case Western Reserve University, as well as the Advanced Research Computing Center's High Performance Computing clusters at the University of Central Florida.
    This work is also part of the CARES project and is funded by the U.S. Department of Energy's Solar Energy Technologies Office under the DE-EE0010418 award.

%
%
\bibliographystyle{splncs04}
\bibliography{refs}
\end{document}